\documentclass[acmsmall,natbib=true,screen,anonymous=False]{acmart}

\usepackage[utf8]{inputenc}

\usepackage{booktabs} 
\usepackage{amsmath}
\usepackage{graphics}
\usepackage{epsfig}
\usepackage{graphicx}
\usepackage{xcolor}
\usepackage{balance}
\usepackage{multirow}
\usepackage{mathrsfs}
\usepackage{acronym}
\usepackage{placeins}
\usepackage{wrapfig}
\usepackage{xcolor}
\usepackage[inline]{enumitem}
\usepackage{fancyhdr}
\usepackage{amsfonts}
\usepackage{svg}
\usepackage{threeparttable}
\usepackage{float}
\usepackage{array}

\usepackage{graphicx}
\usepackage{tabularx}    
\usepackage{booktabs}    
\usepackage{xcolor}      
\usepackage{pifont}      
\usepackage{array}       
\usepackage{enumitem}
\usepackage{multirow} 
\usepackage{amsmath}
\usepackage{subcaption} 
\usepackage[ruled,linesnumbered]{algorithm2e}
\usepackage[noend]{algpseudocode}
\usepackage{xcolor}

\usepackage[most]{tcolorbox} 
\usepackage{fvextra}
\usepackage{colortbl}         

\newcommand{\CheckmarkBold}{\ding{51}}
\newcommand{\XSolidBrush}{\ding{55}}
\SetAlgoSkip{smallskip}   
\SetAlgoInsideSkip{0pt}   

\AtBeginDocument{%
  \providecommand\BibTeX{{%
    \normalfont B\kern-0.5em{\scshape i\kern-0.25em b}\kern-0.8em\TeX}}}

\newcommand{\header}[1]{\vspace*{1mm}\noindent{\textbf{#1}}}

\newcommand{\model}{AdaptSim\xspace}

\DefineVerbatimEnvironment{PromptBlock}{Verbatim}{breaklines=true,breakanywhere=true,fontsize=\scriptsize,formatcom=\ttfamily,frame=single,rulecolor=\color{black!25},framesep=2mm}
\newcounter{prompt}
\renewcommand{\theprompt}{\arabic{prompt}}
\newcommand{\PromptTitle}[2]{%
  \refstepcounter{prompt}%
  \noindent\textbf{Prompt~\theprompt. #1}\label{#2}\par\noindent
}
\newtcolorbox{TranscriptPanel}[2][]{
  enhanced,
  breakable,
  colback=black!2,
  colframe=black!22,
  boxrule=0.45pt,
  arc=1.2mm,
  left=2.5mm,
  right=2.5mm,
  top=1.6mm,
  bottom=1.6mm,
  title={#2},
  fonttitle=\sffamily\bfseries,
  coltitle=black,
  colbacktitle=black!8,
  boxed title style={boxrule=0pt,arc=1.2mm},
  attach boxed title to top left={xshift=2mm,yshift=-1.4mm},
  #1
}

\setcopyright{acmcopyright}
\copyrightyear{2025}
\acmYear{2025} \acmVolume{1} \acmNumber{1} \acmArticle{1} \acmMonth{1}
\acmJournal{TOIS}

\author{Yuanzi Li}
\authornote{Both authors contributed equally to this research.}
\email{liyuanzi0313@outlook.com}
\affiliation{%
  \institution{Renmin University of China}
  \city{Beijing}
  \country{China}
}
\author{Quanyu Dai}
\authornotemark[1]
\email{daiquanyu@huawei.com}
\affiliation{%
  \institution{Huawei Noah’s Ark Lab}
  \city{Shenzhen}
  \country{China}
}

\author{Xueyang Feng}
\email{xueyangfeng@ruc.edu.cn}
\affiliation{%
  \institution{Renmin University of China}
  \city{Beijing}
  \country{China}
}

\author{Zihang Tian}
\email{zihangtian@ruc.edu.cn}
\affiliation{%
  \institution{Renmin University of China}
  \city{Beijing}
  \country{China}
}

\author{Junhao Wang}
\email{2025104106@ruc.edu.cn}
\affiliation{%
  \institution{Renmin University of China}
  \city{Beijing}
  \country{China}
}

\author{Xu Chen}
\authornote{Corresponding author.}
\email{xu.chen@ruc.edu.cn}
\affiliation{%
  \institution{Renmin University of China}
  \city{Beijing}
  \country{China}
}

\author{Zhenhua Dong}
\email{dongzhenhua@huawei.com}
\affiliation{%
  \institution{Huawei Noah’s Ark Lab}
  \city{Shenzhen}
  \country{China}
}

\author{Huifeng Guo}
\email{huifeng.guo@huawei.com}
\affiliation{%
  \institution{Huawei Noah’s Ark Lab}
  \city{Shenzhen}
  \country{China}
}
\renewcommand{\shortauthors}{Yuanzi Li and Quanyu Dai, et al.}

\begin{document}

\title[Towards Fast Domain Adaptation and Fine-Grained User Simulation \\ for Evaluating Conversational Recommender Systems]{Towards Fast Domain Adaptation and Fine-Grained \\ User Simulation for Evaluating \\ Conversational Recommender Systems}

\begin{abstract}
Conversational Recommender Systems (CRSs) enhance user experience through multi-turn interactions, yet accurately evaluating their performance remains challenging.
Recently, employing Large Language Model (LLM) based user simulators for evaluation becomes an effective approach.
However, existing LLM-based user simulators suffer from three key limitations: (1) Lack of Domain Adaptability: The simulators' heavy reliance on fixed prompts and predefined action spaces hinders their ability to adapt and transfer to novel application domains; (2) Limited User Modeling Capability: The simulators lack the capacity for fine-grained user modeling, particularly in accurately replicating subtle linguistic styles and dynamic user preferences; (3) Insufficient Evaluation Validity for Downstream CRSs: When evaluating CRSs, the existing simulators provide not only an unreasonable assessment of fundamental capabilities but also an insufficient examination of system robustness.

To overcome these limitations, we propose \textbf{AdaptSim}, an \textbf{A}daptive \textbf{d}omain and \textbf{a}utomatic \textbf{p}rompt \textbf{t}uning User \textbf{Sim}ulator. AdaptSim offers an efficient and comprehensive framework for evaluating the fundamental capabilities and robustness of CRSs by enabling realistic behavior modeling and diverse style generation.
AdaptSim leverages automatic prompt generation and optimization to reduce manual effort in domain adaptation, and introduces an open action generation mechanism to improve flexibility across domains.
For response generation, we employ controlled text generation combined with a ``think-then-respond'' strategy, enabling fine-grained control over user behavior and language style.
For evaluation of CRSs, AdaptSim incorporates a novel framework centered around a Breadth-First Search (BFS)-based, turn-level pairwise comparison to facilitate a comprehensive and rigorous assessment of the basic capabilities and robustness of CRSs.

Extensive experiments conducted across three domains and with four distinct LLM backbones demonstrate that AdaptSim generates realistic, diverse, and fine-grained dialogues, enabling a more effective and reliable evaluation of the fundamental capabilities and robustness of CRSs.

\end{abstract}


\maketitle
\acresetall

\section{Introduction}
\header{Conversational Recommender Systems.} Conversational Recommender Systems (CRSs) engage users in multi-turn natural language interactions to dynamically capture their preferences and real-time needs, and then provide item recommendations that match their interests~\cite{gao2021advances,jannach2021survey,sun2018conversational}.
They can respond to user queries instantly at any time and significantly reduce manual workloads, thereby playing a crucial role in real-world applications across various domains, such as e-commerce~\cite{liu2023conversational}, online video platforms~\cite{gao2023chat}, and travel services~\cite{liao2019deep}. However, the construction and evaluation of a CRS still face numerous challenges due to its dynamic and interactive nature~\cite{jagerman2019people,gao2021advances}.

\begin{figure}[t]
  \centering
  \includegraphics[width=1\linewidth]{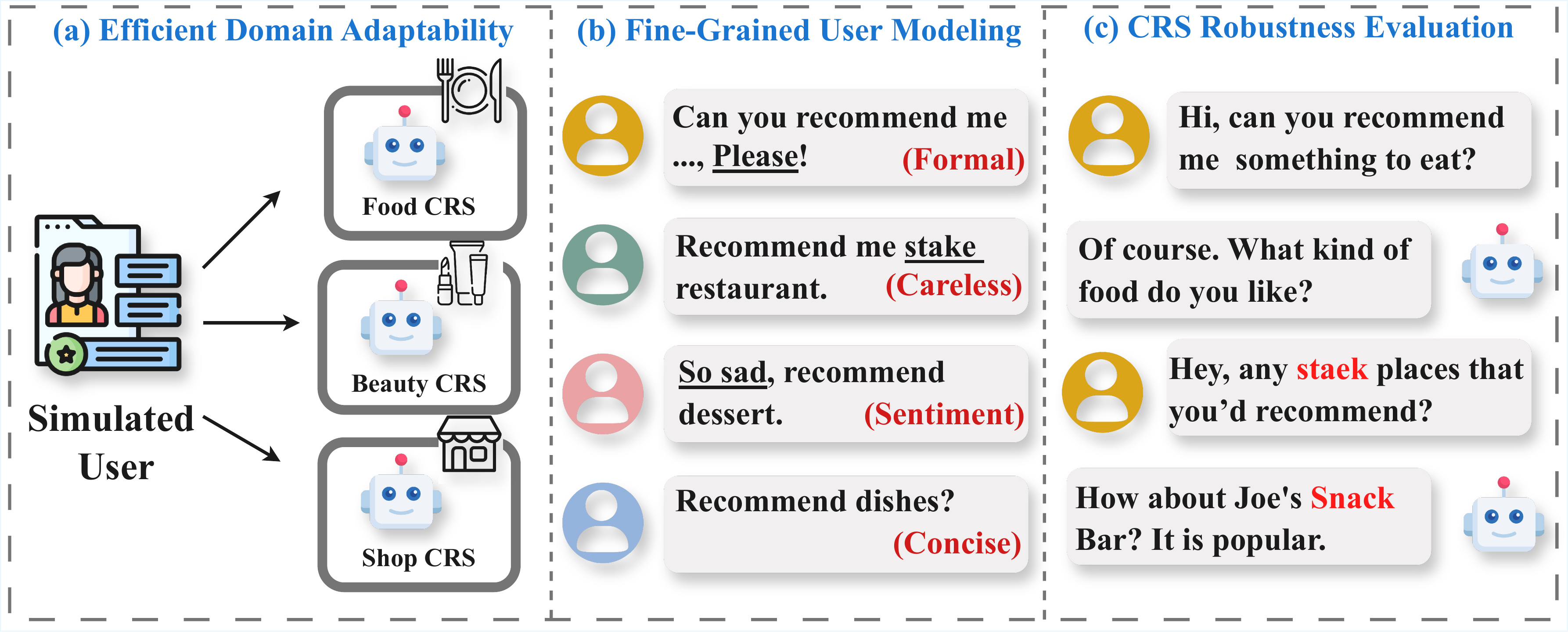}
  \caption{An illustration of the three core capabilities of an ideal user simulator. 
  (a) \textbf{Efficient Domain Adaptability}: The ability to quickly adapt to new domains with minimal human intervention. 
  (b) \textbf{Fine-grained User Modeling}: The capacity to precisely manipulate the stylistic attributes of generated language. 
  (c) \textbf{Sufficient Evaluation Validity}: The capability to provide an accurate, unbiased, and fine-grained evaluation of a CRS's performance and robustness.}
  \label{fig:toy_case}
  \vspace{-1.4em}
\end{figure}

\header{User Simulator for CRS.} To address this issue, user simulators have emerged as promising solutions recently~\cite{lei2020interactive,sun2018conversational,zhang2020evaluating}. 
To effectively support diverse application scenarios, a user simulator should possess three key characteristics as illustrated in Figure~\ref{fig:toy_case}: 
(1) \emph{Efficient Domain Adaptability}. It enables the simulator to adapt to different application domains efficiently. 
(2) \emph{Fine-grained User Modeling}. It allows the simulator to achieve realistic role-playing and accurate performance evaluation.
(3) \emph{Sufficient Evaluation Validity}. It ensures the simulator provides a comprehensive and rigorous assessment of the basic capabilities and robustness of CRSs.
The recent remarkable performance of large language models (LLMs) in dialogue comprehension, text generation, and role-playing has driven the rapid development of LLM-based user simulations.
These techniques encompass diverse paradigms, ranging from single-prompt methods~\cite{wang2023rethinking,zhu2024reliable,fang2024multi} to agent-based architectures~\cite{chen2025recusersim,zhu2025llm}.

\header{Remaining Limitations.} Although current user simulation approaches have made significant progress, they still face several critical limitations:

(1) \emph{Lack of Domain Adaptability}. These methods typically rely on manually crafted fixed prompt templates~\cite{fang2024multi} and a predefined, limited action space~\cite{wang2023rethinking,chen2025recusersim,zhu2025llm}, both of which often require careful design and maintenance by domain experts. This not only incurs high development costs but also significantly restricts the simulator’s domain adaptation capability.
Consequently, they lack the flexibility and scalability to efficiently adapt to new domains or diverse dialogue contexts in real-world applications.

(2) \emph{Limited User Modeling Capability.} 
Existing methods~\cite{wang2023rethinking,chen2025recusersim,zhu2025llm,fang2024multi} generally lack fine-grained modeling of users' linguistic behavior.
The generated responses often remain at a superficial semantic level, failing to capture deeper linguistic attributes such as style, emotion, and politeness strategies, thereby undermining the realism and interactivity of simulated dialogues.
One attempt~\cite{chen2025recusersim} aims to improve language personalization through style control mechanisms. However, its reliance on refine-based post-processing frameworks disrupts contextual coherence, leading to character inconsistency and illogical transitions.

(3) \emph{Insufficient Evaluation Validity.} 
Evaluating CRSs with ground-truth dialogues~\cite{wang2023rethinking,fang2024multi,zhou2020improving} is often impractical due to data scarcity. The limitation led to a shift towards using LLMs to simulate user profiles and generate dialogues with CRS, followed by dialogue-level pointwise scoring to assess the CRS.~\cite{zhu2024llmbasedcontrollablescalablehumaninvolved,chen2025recusersim}. However, this new paradigm introduces its own critical flaws. Its pointwise scoring methods are prone to ``Flattery Bias''~\cite{zheng2023judging,chang2024survey} and ``Rough Scoring Bias''~\cite{li2025evaluating,tripathi2025pairwise}, while its coarse, dialogue-level scope misses turn-level errors. Furthermore, simulators built on powerful backbones create unrealistically perfect dialogues. This establishes a reality gap that inflates performance metrics for CRSs and fails to test their robustness against noisy, imperfect user behavior.

\header{Our Proposed Method.} To address the above limitations, we propose AdaptSim, a high-quality user simulator capable of rapid domain adaptation and the generation of diverse linguistic styles. As shown in Table~\ref{tab:related}, AdaptSim enables comprehensive evaluation of the basic capabilities of CRS and its robustness under varying stylistic conditions. Specifically, AdaptSim addresses the aforementioned three challenges through the following four key mechanisms: 

(1) \emph{Automatic Prompt Optimization.} By leveraging domain-specific knowledge and historical interaction data, it automatically refines prompts to enable the simulator to generate high-quality dialogues across all domains. 

(2) \emph{Open-ended Action Generation.} AdaptSim moves away from relying on a set of predefined actions and instead adopts an open-ended action generation mechanism, dynamically constructing user actions based on context, which significantly boosts the simulator's flexibility across new domains. 

(3) \emph{Thinking-Enhanced Style Generation.} To enhance fine-grained control over linguistic behavior of users, we adopt a ``think-then-respond'' structure in response generation. This method guides the model to think on its language style before generating a response, significantly improving stylistic consistency while maintaining contextual coherence.

(4) \emph{BFS-based Turn-level Pairwise Evaluation.} To achieve an unbiased and fine-grained turn-level assessment of a CRS's basic capabilities and robustness, this method combines three key elements. It leverages a Breadth-First Search (BFS) to ensure contextual consistency, while the turn-level analysis provides detailed feedback, and the pairwise comparison framework guarantees objectivity.

\header{Contributions.} Our main contributions are summarized as follows:
\begin{itemize}[label=\textbullet,leftmargin=*]
    \item We design automated prompt optimization, open-ended action generation, and a ``think-then-respond'' architecture to enable rapid cross-domain adaptation and fine-grained control over simulator language.
    
    \item We introduce a BFS-based turn-level pairwise evaluation to perform an accurate, unbiased, and fine-grained assessment of a CRS's basic capabilities and robustness. 

    \item Experiments show that, using the generated prompts, our method outperforms existing approaches across multiple dimensions and achieves significant results in language control. Moreover, the CRS evaluation results based on our method further validate its effectiveness.
\end{itemize}

\begin{table*}[t]
    \centering
    \caption{Related work on user simulators in CRS. For the simulators themselves, ``Style'', and ``Adaptability'' refer to, language style, and fast adaptation across domains, respectively. For CRS evaluation, ``Basics'' refers to the ability to perform unbiased evaluation of the CRS basics, while ``Robustness'' refers to the ability to assess the CRS's robustness.
    }
    \begin{tabularx}{0.8\textwidth}{c *{5}{>{\centering\arraybackslash}X}} 
      \toprule
      & \multicolumn{2}{c}{Simulator} & \multicolumn{2}{c}{CRS\_Test} \\
      \cmidrule(lr){2-3}\cmidrule(lr){3-5}
      Name & Style & Adaptability & Basics & Robustness \\
      \midrule
      iEvaLM~\cite{wang2023rethinking} & \textcolor{red}{\XSolidBrush} & \textcolor{red}{\XSolidBrush} &\textcolor{cyan}{\CheckmarkBold} & \textcolor{red}{\XSolidBrush} \\
      SimpleUserSim~\cite{zhu2024reliable} 
     & \textcolor{red}{\XSolidBrush} & \textcolor{red}{\XSolidBrush} &\textcolor{cyan}{\CheckmarkBold} & \textcolor{red}{\XSolidBrush} \\
      CSHI~\cite{zhu2025llm} 
      & \textcolor{red}{\XSolidBrush} & \textcolor{red}{\XSolidBrush} & \textcolor{red}{\XSolidBrush} & \textcolor{red}{\XSolidBrush} \\
      RecUserSim~\cite{chen2025recusersim} 
      &\textcolor{cyan}{\CheckmarkBold} & \textcolor{red}{\XSolidBrush} & \textcolor{red}{\XSolidBrush} & \textcolor{red}{\XSolidBrush} \\
      Ours 
       & \textcolor{cyan}{\CheckmarkBold} & \textcolor{cyan}{\CheckmarkBold} & \textcolor{cyan}{\CheckmarkBold} & \textcolor{cyan}{\CheckmarkBold} \\
      \bottomrule
    \end{tabularx}
    \label{tab:related}
    \vspace{-0.5em}
  \end{table*}

\section{Related Work}
In this section, we provide a literature review of the research about conversational recommendation systems, LLM based user simulator in CRS and automatic prompt optimization.

\subsection{Conversational Recommendation Systems} Conversational recommender systems (CRSs) aim to support recommendation through interactive natural language exchanges, where the system gradually elicits user preferences and provides personalized suggestions~\citep{zhang2024towards}. Existing CRS methods are commonly grouped into attribute-based and generation-based paradigms~\citep{CRS_Survey_Jannach}. Attribute-based systems usually guide conversations through predefined slots or templates, which makes their behavior easier to control but limits conversational flexibility. Generation-based systems relax these constraints and produce more natural responses~\citep{Wang_2022}, yet traditional generation models often struggle with long-context understanding, preference tracking, and robust dialogue planning.

Recent work has explored LLMs as a new foundation for CRSs. ZSCRS~\citep{he2023zscrs} studies the zero-shot recommendation ability of LLMs in conversational settings, while ChatRec~\citep{gao2023chatrec} and InteRecAgent~\citep{huang2023interecagent} combine LLMs with external recommendation models or tools to improve interactive recommendation. To further strengthen dialogue management, MACRS~\citep{fang2024multi} introduces a multi-agent framework for strategic planning in multi-turn recommendation. These studies show the potential of LLM-based CRSs, but they also make evaluation more challenging: system responses become more open-ended, dialogue trajectories diverge quickly, and performance depends on both recommendation quality and conversational behavior. Therefore, reliable user simulation and fine-grained evaluation protocols are increasingly important for assessing CRS capabilities under realistic interaction conditions.

\subsection{LLM Based User Simulator in CRS}
In the study of user simulation in CRS leveraging Large Language Models, existing methods can be broadly categorized into two main paradigms: single-prompt frameworks and agent-based frameworks.

Early works, such as iEvaLM \cite{wang2023rethinking}, MACRS \cite{fang2024multi}, and SimpleUserSim \cite{fang2024multi}, typically adopt the single-prompt paradigm. These approaches rely on manually designed static templates to consolidate user history, candidate items, and task instructions into a single prompt, thereby generating a simulated user response in a single pass. However, the inherent rigidity of this approach leads to several limitations. First, because the generation process is heavily constrained by a fixed template structure, the resulting dialogue flows exhibit limited diversity and often fall into predictable and repetitive patterns. Second, this one-shot generation mechanism makes fine-grained control over the response content challenging, hindering the ability to dynamically adjust the simulated user's intent or emotional state. Finally, these methods typically rely on ground-truth labels to evaluate the dialogue system, which places emphasis solely on the system’s recommendation accuracy while overlooking its conversational competence. Moreover, ground-truth data are scarce in most real-world scenarios, further constraining the applicability and scalability of such approaches.

To overcome the aforementioned limitations, researchers have shifted towards more dynamic, agent-based frameworks. The LLM-driven user agent, CSHI \cite{zhu2025llm}, represents a notable attempt in this direction. It guides the agent's behavior by explicitly integrating user profiles and a fixed action space. This design aims to enhance the reliability and personalization of simulated behaviors, making the agent's decisions more grounded. Despite this advancement, CSHI still faces challenges in consistency, behavioral diversity, and persona stability in practice.
Building upon CSHI, RecUserSim \cite{chen2025recusersim} further expands CSHI's capabilities by augmenting its action space and modeling linguistic styles to improve the diversity of simulated user behavior. However, its core design does not escape the constraints of a predefined framework. Although its action space is more extensive, it remains fundamentally fixed and fails to capture the open-ended and unanticipated behaviors humans exhibit in real interactions. Furthermore, the post-refinement strategy employed by RecUserSim, while capable of locally optimizing response quality, sometimes compromises contextual coherence. This can disrupt the natural flow of the dialogue, thereby limiting its overall performance.

In summary, both the single-prompt and the existing agent-based frameworks suffer from a common, fundamental bottleneck: a heavy reliance on manually crafted prompts and predefined action spaces. This strong dependency not only demands significant expert knowledge and engineering effort but, more critically, it severely hinders the simulator's adaptability to new domains or dynamic environments.

\begin{figure*}[t]
  \centering
  \includegraphics[width=1\linewidth]{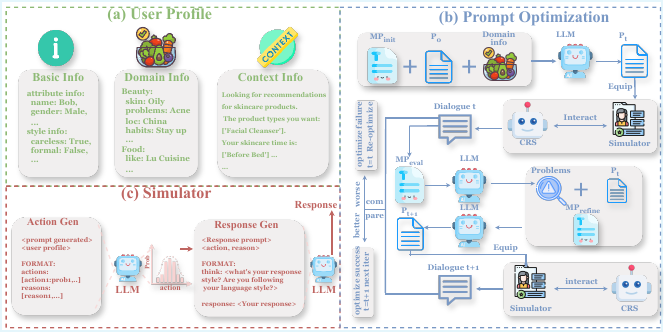}
\caption{\textcolor[HTML]{B2B366}{(a) User Profile} Shows the multi-domain profiles, demonstrating that our simulator can quickly adapt to new domains. \textcolor[HTML]{6C8EBF}{(b) Prompt Optimization} Illustrates the automatic prompt optimization process when adapting to a new domain: first, domain adaptation is performed, followed by iterative optimization of the prompt based on interactions with CRS. \textcolor[HTML]{B85450}{(c) Simulator.} After optimizing the prompt, we integrate it into the simulator and employ an open-ended action generation method to enhance domain adaptability, along with the ``think-then-respond'' approach to control language style.}
  \label{fig:pipeline}
  \vspace{-0.8em}
\end{figure*}
\subsection{Automatic Prompt Optimization}

Automatic Prompt Optimization (APO) is an emerging technique aimed at automatically discovering and refining input prompts for language models through iterative feedback and evaluation. Its core objective is to replace the tedious and often intuitive process of manual prompt engineering. Existing approaches have explored various strategies in this domain. For instance, methods like APE \cite{zhou2022large} iteratively search for better instructions in a propose-and-score paradigm, combining manually designed seed prompts with feedback from the model itself. Other approaches, such as GPS \cite{xu2022gps} and SPRIG \cite{zhang2024sprig}, integrate human feedback into the optimization loop, leveraging human preferences to guide the evolution of prompts to enhance their quality and alignment. Meanwhile, to reduce the reliance on human intervention, methods like DAPO \cite{yang2024dual} attempt to leverage the language model's own capabilities for instruction induction, automatically inferring the optimal task instruction from a few examples.

Although these methods have achieved promising results in static tasks (e.g., text classification, summarization), they face significant challenges when applied to dynamic, multi-turn user simulation scenarios. Static tasks typically have clear, fixed evaluation criteria, such as accuracy or overlap with a reference answer, whereas the quality assessment of multi-turn dialogues is considerably more complex. The core difficulty lies in the lack of an objective, quantifiable comparative standard across turns. In a simulated dialogue, a ``good'' response must not only be locally optimal for the current turn but also maintain consistency and coherence with the historical context, while steering the conversation in a meaningful direction. Existing APO methods, typically designed to optimize for single-turn interactions, struggle to capture this long-term dependency on global conversational quality.

To address these issues, we draw inspiration from the work of ProTeGi \cite{pryzant2023automatic}, which demonstrates the effectiveness of optimizing generation strategies via pairwise preference learning in an unsupervised setting. We apply this concept to prompt optimization for dynamic multi-turn user simulation by proposing a novel approach. This approach introduces a multi-round interaction mechanism, where the model generates entire conversational snippets rather than single responses, combined with pairwise comparisons of historical outputs for evaluation. Specifically, the system generates two conversational trajectories driven by different candidate prompts, and then a model or a human annotator determines which trajectory is globally superior (e.g., more coherent or more aligned with the user persona). This design allows the prompt refinement process to move beyond a reliance on hard-to-obtain ground-truth labels, instead leveraging relative preferences for iterative optimization. Consequently, our method not only effectively improves prompt quality for dynamic interaction tasks but also significantly reduces supervision costs, offering a viable path for optimizing user simulators in complex scenarios.

\section{Preliminaries}

In this section, we survey the evaluation landscape for CRS. We begin by reviewing the established methodologies for assessing a system's basic capabilities. We then shift our focus to robustness testing—a comparatively less explored but critical area—to define its scope and outline corresponding evaluation approaches.

\subsection{Evaluation of CRS Basic Capabilities}
Previous methodologies for evaluating the basic capabilities of a CRS have primarily relied on three comparison strategies, each with significant drawbacks:

\begin{enumerate}[leftmargin=*]
    \item \textbf{Dialogue-level Pointwise Comparison}: This approach assigns a single score to an entire conversation. It is often too coarse-grained, as one score can mask specific turn-level failures and thus lacks diagnostic detail. 

    \item \textbf{Turn-level Pointwise Comparison}: While more granular, this method evaluates each turn in relative isolation. This can lead to context-agnostic judgments that fail to assess the dialogue's overall coherence or long-term strategy. Furthermore, the use of discrete rating scales (e.g., 1-5) can lead to quantization errors, while LLM-based evaluators may introduce a positivity bias that obscures meaningful differences between systems.
    
    \item \textbf{Dialogue-level Pairwise Comparison}: This method compares the complete dialogues from two different CRS agents to select a winner. However, as the conversations diverge, the contexts at each turn can become substantially different, making a direct comparison in later stages unfair and unreliable.
\end{enumerate}

\subsection{CRS Robustness}

CRS robustness refers to the system's ability to maintain performance and accurately infer user intent when faced with diverse and imperfect user inputs. In practice, users often generate non-standard inputs due to spelling errors, ambiguous language, colloquial expressions, or incomplete information. A truly robust CRS must be able to effectively recognize and process such deviations to fulfill the user's request, rather than misunderstanding or failing completely.

To illustrate a common failure in robustness, we constructed a simple test case. As shown in Figure~\ref{fig:toy_case}, a user simulator intentionally introduces a common spelling error—typing ``staek'' instead of ``steak''—to mimic a realistic user mistake. Despite this minor input deviation, the CRS misunderstands the user's intent and provides an irrelevant recommendation: ``Joe's Snack Bar.'' This example highlights a significant vulnerability in current systems when handling the inherent uncertainty of real-world user language.

\section{Method}
To address the challenges of poor domain adaptability and limited user modeling capability, we propose \textbf{AdaptSim}, an LLM-based \underline{\textbf{A}}daptive \underline{\textbf{d}}omain and \underline{\textbf{a}}utomatic \underline{\textbf{p}}rompt \underline{\textbf{t}}uning User \underline{\textbf{Sim}}ulator. 
As illustrated in Figure~\ref{fig:pipeline}, AdaptSim first employs automatic prompt optimization to achieve cross-domain adaptation efficiently (Section~\ref{sec:APO}).
It then introduces a two-level control mechanism to enable realistic user modeling: at the strategic level, we design an open-ended action generator for dialogue strategy planning (Section~\ref{sec:openend}); at the linguistic level, we develop a style controller based on a ``think-then-respond'' mechanism to enable fine-grained language controllability (Section~\ref{sec:thinkenhanced}).

\subsection{Automatic Prompt Optimization via Simulator-CRS Feedback}\label{sec:APO}
Existing simulator methods typically rely on manually crafted prompts, which not only lack domain adaptability but also require extensive expert knowledge for design and tuning. To address this, we propose an automated prompt optimization framework.
Rather than treating prompt construction as a static design problem, we formulate it as an iterative optimization process composed of two main stages:
(1) domain adaptation via initialization, and
(2) iterative refinement based on feedback
with termination triggered by a convergence condition.

\begin{figure}[t]
   \begin{algorithm}[H]
    \SetAlgoLined
    \KwIn{Max iterations $\mathrm{T}$, Patience threshold $\mathrm{P}$, Initial $\mathrm{Prompt_0}$, Domain info $\mathrm{D}$}
    \KwOut{Best performing prompt}
    
    $\mathrm{noCount} \gets 0$; $\mathrm{t} \gets 1$\;
    \vspace{3pt}
    \textbf{\textcolor{gray}{// Domain Adaptation Phase}}\;
    $\mathrm{Prompt_1} \gets \mathrm{LLM}(\mathrm{MP}_{\text{init}}(\mathrm{D}, \mathrm{Prompt_0}))$\;
    $B_1 \gets \mathrm{AdaptSim}(\mathrm{Prompt_1})$\;
    \vspace{3pt}
    
    \textbf{\textcolor{gray}{// Iterative Refinement Phase}}\;
    \While{$\mathrm{t} < \mathrm{T}$ \textnormal{and} $\mathrm{noCount} < \mathrm{P}$}{
        $\mathrm{E_t} \gets \emptyset$\;
        \vspace{3pt}
        
        \textbf{\textcolor{gray}{// Error Detection}}\;
        \ForEach{$\mathrm{d_k^{(t)}} \in \mathrm{B_t}$}{
            $\mathrm{F_{k}^{(t)}} \gets \mathrm{LLM}(\mathrm{MP}_{\text{eval}}(\mathrm{d_k^{(t)}}))$\;
            $\mathrm{E_t} \gets \mathrm{E_t} \cup \{\mathrm{F_{k}^{(t)}\}}$\;
        }
        \vspace{3pt}
        
        \textbf{\textcolor{gray}{// Prompt Refinement}}\;
        $\mathrm{Prompt_{t+1}} \gets \mathrm{LLM}(\mathrm{MP}_{\text{refine}}(E_t), \mathrm{Prompt_t)}$\;
        $\mathrm{B_{t+1}} \gets \mathrm{AdaptSim}(\mathrm{Prompt_{t+1}})$\;
        \vspace{3pt}
        
        \textbf{\textcolor{gray}{// Stopping Criteria Changed}}\;
        $\mathrm{noCount} \gets (\mathrm{quality}(\mathrm{B_{t+1}}) > \mathrm{quality}(\mathrm{B_t})) \ ? \ 0 \ : \ \mathrm{noCount} + 1$\;
        $\mathrm{t} \gets \mathrm{t + 1}$\;
    }
    \caption{APO via Simulator-CRS Feedback}
    \label{alg:prompt_optimization}
\end{algorithm} 
\vspace{-1em}
\end{figure}

\subsubsection{Iteration 0: Domain Adaptation}
The process begins by constructing the initial domain-specific prompt $Prompt_1$ as follows:
\begin{equation}
\mathrm{Prompt_1} = \mathrm{LLM}\left( \mathrm{MetaPrompt_{init}}(\mathcal{\mathrm{D}}, \mathrm{Prompt_0}) \right)
\end{equation}
where $\mathcal{D}$ denotes the domain-specific context and $Prompt_0$ is a general-purpose template shared across domains. The initial prompt $Prompt_1$ is used in conjunction with AdaptSim to simulate a batch of synthetic dialogues:
\begin{equation}
\mathrm{B_1} = \{\mathrm{d_1^{(1)}, d_2^{(1)}, \ldots, d_N^{(1)}}\}  
\end{equation}
This phase provides domain adaptation through data-driven prompt construction without manual effort.

\subsubsection{Iterations 1-n: Error Detection and Prompt Refinement}
In each iteration, the dialogue batch $B_{t}$ is evaluated to identify behavioral or stylistic issues, which are aggregated into a problem set:
\begin{equation}
\mathrm{E_t = \bigcup_{k=1}^{N} F_k^{(t)} } 
\end{equation}
where \( \mathrm{F_k^{(t)}} \) represents the set of issues identified from the dialogue \( \mathrm{d_k^{(t)}} \) in the batch using a meta-evaluation prompt:
\begin{equation}
\mathrm{F_k^{(t)} = \mathrm{LLM}\left( \mathrm{MetaPrompt_{eval}}\left( d_k^{(t)} \right) \right)}
\end{equation}
We analyze the problem set $E_{t}$ to identify recurring failure patterns such as underspecified intent or inconsistent guidance. Based on this analysis, an improved prompt is generated:
\begin{equation}\mathrm{
Prompt_{t+1} = \mathrm{LLM}\left( \mathrm{MetaPrompt}_{refine}(E_{t}, Prompt_t )\right)  }
\end{equation}
Subsequently, the updated prompt is used to generate a new batch of data, $\mathrm{B_{t+1}}$. To concretely evaluate its quality improvement, we introduce an LLM as a judge to perform pairwise comparisons between samples from $B_{t+1}$ and $\mathrm{B_t}$. We define the criterion for improvement as follows: the quality of $\mathrm{B_{t+1}}$ is considered superior to that of $\mathrm{B_t}$ if its win rate is greater than the loss rate, i.e., $\mathrm{\text{win\_rate}(B_{t+1} \text{ vs } B_t) > \text{loss\_rate}(B_{t+1} \text{ vs } B_t)}$.

\subsubsection{Stopping Criteria}
The iterative optimization proceeds until one of the following conditions is satisfied:
\begin{itemize}[label=\textbullet,leftmargin=*]
    \item A predefined maximum number of iterations is reached;
    \item No substantial performance improvement is observed over three consecutive iterations.
\end{itemize}

The detailed algorithmic procedure is summarized in Algorithm~\ref{alg:prompt_optimization}, and the full list of meta-prompts is provided in Prompts~\ref{tab:prompt1}--\ref{tab:prompt6}.

Through the aforementioned APO process, we generate domain-adapted prompts. In section~\ref{sec:openend} and~\ref{sec:thinkenhanced}, we will apply both Open-ended Action Generation and Thinking-Enhanced Style Generation operations on this prompt to achieve two-level control mechanism.

\subsection{Strategy-Level Control: Open-Ended Action Generation}\label{sec:openend}
To enable dialogue strategy planning in user simulation, AdaptSim integrates a strategy-level control mechanism that directly generates user actions instead of selecting from a predefined action set. This approach enhances realism and flexibility, particularly in open-domain scenarios where user behavior may not adhere to a rigid structure.
The core idea is inspired by the Bayesian Brain Hypothesis~\cite{KNILL2004712}, which posits that the human brain functions as a probabilistic inference engine during decision-making. Following this, we model action generation as a probabilistic reasoning process: the simulator generates multiple candidate actions, estimates their likelihoods, and selects the one with the highest probability as the user’s next move.

Formally, given the dialogue history $\mathcal{H}$ and user profile $\mathcal{P}$, we treat action generation as probabilistic inference over candidate actions. Conditioned on the prompt constructed from $\mathcal{H}$ and $\mathcal{P}$, the LLM induces an action distribution:
\begin{equation}
\pi(a \mid \mathcal{H}, \mathcal{P}) = \mathrm{LLM}\left( \mathrm{Prompt}_{\mathrm{action}}(\mathcal{H}, \mathcal{P}) \right),
\end{equation}
where $\pi(a \mid \mathcal{H}, \mathcal{P})$ denotes the probability of candidate action $a$ conditioned on the dialogue history and user profile. The simulator then selects the most probable action:

\begin{equation}
a^\ast = \arg\max_{a} \ \pi(a \mid \mathcal{H}, \mathcal{P})
\end{equation}
where $a^\ast$ denotes the selected action. This formulation enables AdaptSim to generate open-ended actions while remaining grounded in the current dialogue context and user profile.

\subsection{Language-Level Control: Thinking-Enhanced Style Generation}\label{sec:thinkenhanced}
Beyond action strategy planning, realistic user simulation also requires fine-grained control over linguistic expression. 
In multi-turn dialogues, the accumulation of contextual noise and the weakening of style objectives often cause traditional controlled text generation methods to deviate from the intended stylistic constraints~\cite{liang2024controllabletextgenerationlarge}.
To address this issue, we draw inspiration from the Metacognition Theory~\cite{Metacognitive}, which emphasizes the role of self-monitoring and self-regulation in cognitive processes. Based on this theory, we design a language-level control mechanism that introduces an explicit \textit{thinking step} prior to response generation. This step allows the simulator to reflect on its stylistic goals and prepare for consistent expression. Specifically, before generating the final response, the model is required to complete a structured reflection step in the following format:

\begin{quote}
\texttt{<think>: What's your response style? Are you following your  language style?}
\end{quote}

This self-assessment step encourages the model to evaluate its stylistic intention and adherence prior to generation, effectively enhancing the controllability and consistency of the generated text. The final response is then generated as:
\begin{equation}
    R = \mathrm{LLM}\left( \mathrm{Prompt_{response}}(\text{S}, a^\ast, \mathcal{H}, \mathcal{P}) \right)
\end{equation}

where $S$ denotes the generated language style.
By simulating a ``think-then-respond'' process similar to human communication, this mechanism significantly improves the controllability of linguistic behavior.

\subsection{BFS-Based Turn-Level Pairwise Evaluation}
To evaluate the basic capabilities and robustness of conversational recommender systems (CRSs), we introduce a BFS-based turn-level pairwise evaluation protocol. The goal is to compare systems under matched dialogue contexts rather than independently evolving conversations. By expanding dialogue states with Breadth-First Search (BFS), the protocol preserves context consistency at each comparison point, while turn-level pairwise judgments provide fine-grained signals and reduce the instability of absolute scoring.

\subsubsection{BFS-Based Turn-Level Pairwise Evaluation for CRS Basics}
\label{sec:BFS_Bas}

As outlined in Algorithm~\ref{alg:BFS1_revised}, we use a tree-structured simulation process to evaluate the basic capabilities of CRSs. Unlike linear simulation, where two systems quickly enter different dialogue contexts after producing different responses, our BFS-based protocol maintains local \textit{context consistency} for each pairwise comparison.

The process starts from an initial user query $a_0$ and initializes a queue $\mathrm{Q}$ with the corresponding dialogue history. At each iteration, the algorithm dequeues a history state $h$ and asks both target systems, $\mathrm{CRS}_1$ and $\mathrm{CRS}_2$, to generate responses $r_1$ and $r_2$ from the same history. This synchronous generation step ensures that $r_1$ and $r_2$ are evaluated under an identical context, avoiding bias caused by context divergence.

The user simulator $\mathrm{U}$ then interacts with the two responses independently, producing two branches: $h \cdot r_1 \cdot a_1$ and $h \cdot r_2 \cdot a_2$. These new states are added back to the queue, allowing the evaluation to expand breadth-wise over multiple possible dialogue trajectories. A branch terminates when the simulator outputs ``END'' or when the maximum depth $\mathrm{R}$ is reached.

\subsubsection{BFS-Based Turn-Level Pairwise Evaluation for CRS Robustness}
\label{sec:BFS_Rob}

As shown in Algorithm~\ref{alg:BFS2}, we adapt the same tree-structured protocol to evaluate CRS robustness under user-side noise. Instead of comparing two systems, this setting performs a \textit{perturbation-based comparison} within a CRS. We use two user simulators: a careless user $\mathrm{U}_c$, which produces noisy inputs such as typos or incomplete expressions, and a normal user $\mathrm{U}_n$, which produces clean inputs.

For each dequeued history state $h$, the two simulators generate a noisy query $a_c$ and a clean query $a_n$ in parallel. The target CRS responds to them separately, producing $r_c$ and $r_n$. Since both inputs are generated from the same dialogue history, the resulting pairwise comparison isolates the effect of input noise and measures how much the CRS performance changes under perturbation.

To avoid cumulative error propagation, we use a \textit{clean-context preservation} strategy during tree expansion. The noisy branch $h \cdot a_c \cdot r_c$ is used only for evaluation and is not enqueued. Instead, only the clean trajectory $h \cdot a_n \cdot r_n$ is retained for subsequent turns. This design keeps later comparisons grounded in coherent dialogue histories while still measuring the immediate effect of noisy user input at each turn.

\begin{figure}[t]
   \begin{algorithm}[H]
    \SetAlgoLined
    \caption{Dialogue Simulation using BFS to Fairly Evaluate CRSs' Basic capabilities}
    \label{alg:BFS1_revised}
    \KwIn{User simulator $\mathrm{U}$, Dialogue systems $\mathrm{CRS}_1, \mathrm{CRS}_2$, Max rounds $\mathrm{R}$}
    \KwOut{Pairwise evaluation results}
    
    \textbf{\textcolor{gray}{// Initialization}}\;
    $\mathrm{a}_0 \gets \mathrm{U}.\mathrm{ask}()$\;
    Initialize queue $\mathrm{Q} \gets [(\mathrm{a}_0, 0)]$\;
    
    \textbf{\textcolor{gray}{// Breadth-First Search Simulation}}\;
    \While{$\mathrm{Q}$ is not empty}{
        $(\mathrm{h}, \mathrm{t}) \gets \mathrm{Dequeue}(\mathrm{Q})$\;
        \If{$\mathrm{t} \geq \mathrm{R}$}{
            \textbf{continue}\;
        }
        
        \textbf{\textcolor{gray}{// Generate system replies based on current history}}\;
        $\mathrm{r}_1 \gets \mathrm{CRS}_1.\mathrm{reply}(\mathrm{h})$\;
        $\mathrm{r}_2 \gets \mathrm{CRS}_2.\mathrm{reply}(\mathrm{h})$\;
        
        Perform pairwise evaluation of $\mathrm{r}_1$ and $\mathrm{r}_2$\;
        
        \textbf{\textcolor{gray}{// Generate user responses and expand search tree}}\;
        $\mathrm{a}_1 \gets \mathrm{U}.\mathrm{respond}(\mathrm{h}, \mathrm{r}_1)$\;
        \If{$\mathrm{a}_1 \neq$ ``END''}{
            $\mathrm{Enqueue}(\mathrm{Q}, (\mathrm{h} \cdot \mathrm{r}_1 \cdot \mathrm{a}_1, \mathrm{t}+1))$\;
        }
        
        $\mathrm{a}_2 \gets \mathrm{U}.\mathrm{respond}(\mathrm{h}, \mathrm{r}_2)$\;
        \If{$\mathrm{a}_2 \neq$ ``END''}{
            $\mathrm{Enqueue}(\mathrm{Q}, (\mathrm{h} \cdot \mathrm{r}_2 \cdot \mathrm{a}_2, \mathrm{t}+1))$\;
        }
    }
\end{algorithm}
\vspace{-1em}
\end{figure}

\begin{figure}[t]

\begin{algorithm}[H]
    \SetAlgoLined
    \caption{Dialogue Simulation using BFS to Evaluate CRS Robustness}
    \label{alg:BFS2}
    \KwIn{Careless user simulator $\mathrm{U}_c$, Normal user simulator $\mathrm{U}_n$, Dialogue system $\mathrm{CRS}$, Max rounds $\mathrm{R}$}
    \KwOut{Pairwise evaluation results on robustness}
    
    \textbf{\textcolor{gray}{// Initialization}}\;
    $\mathrm{a}_0 \gets \mathrm{U}_n.\texttt{ask}()$\;
    $\mathrm{r}_0 \gets \mathrm{CRS}.\texttt{reply}(\mathrm{a}_0)$\;
    Initialize queue $\mathrm{Q} \gets [(\mathrm{a}_0 \cdot \mathrm{r}_0, 0)]$\;
    
    \textbf{\textcolor{gray}{// Breadth-First Search Simulation}}\;
    \While{$\mathrm{Q}$ is not empty}{
        $(\mathrm{h}, \mathrm{t}) \gets \mathrm{Dequeue}(\mathrm{Q})$\;
        \If{$\mathrm{t} \geq \mathrm{R}$}{
            \textbf{continue}\;
        }
        
        \textbf{\textcolor{gray}{// Generate responses from different user simulators based on history}}\;
        $\mathrm{a}_c \gets \mathrm{U}_c.\texttt{respond}(\mathrm{h})$,
        $\mathrm{a}_n \gets \mathrm{U}_n.\texttt{respond}(\mathrm{h})$\;
        
        \textbf{\textcolor{gray}{// Generate system replies to different user responses}}\;
        $\mathrm{r}_c \gets \mathrm{CRS}.\texttt{reply}(\mathrm{h} \cdot \mathrm{a}_c)$,
        $\mathrm{r}_n \gets \mathrm{CRS}.\texttt{reply}(\mathrm{h} \cdot \mathrm{a}_n)$\;
        
        Perform pairwise evaluation of $\mathrm{r}_c$ and $\mathrm{r}_n$\;
        
        \textbf{\textcolor{gray}{// Expand search tree with both user behaviors}}\;
        \If{$\mathrm{a}_n \neq$ ``END''}{
            $\mathrm{Enqueue}(\mathrm{Q}, (\mathrm{h} \cdot \mathrm{a}_n \cdot \mathrm{r}_n, \mathrm{t}+1))$\;
        }
    }
\end{algorithm}
\vspace{-1em}
\end{figure}

\section{Experiments}
In this work, we investigate two groups of research questions. The first evaluates the effectiveness of \model\ as a user simulator, while the second examines whether \model\ can support reliable evaluation of conversational recommender systems (CRSs). Accordingly, we organize our research questions as follows:

  \begin{itemize}[label=$\circ$,leftmargin=*]
  \item \textbf{RQ1: Effectiveness and generalizability of \model\ as a user simulator}
  \begin{itemize}[label=\textbullet,leftmargin=*]
      \item \textbf{RQ1.1:} Does \model\ outperform existing user simulators in simulation quality?
      \item \textbf{RQ1.2:} Can \model\ provide fine-grained control over users' linguistic styles?
      \item \textbf{RQ1.3:} Can \model\ generalize to more open-ended dialogue scenarios, such as chit-chat systems?
      \item \textbf{RQ1.4:} Does each component of \model\ contribute effectively to overall performance?
      \item \textbf{RQ1.5:} How does \model\ address common issues observed in existing simulators through case studies?
  \end{itemize}

  \item \textbf{RQ2: Reliable CRS evaluation with \model}
  \begin{itemize}[label=\textbullet,leftmargin=*]
      \item \textbf{RQ2.1:} Can \model\ enable reliable evaluation of the core capabilities of CRSs?
      \item \textbf{RQ2.2:} Can \model\ enable reliable evaluation of CRS robustness?
  \end{itemize}
  \end{itemize}

\subsection{Experiments Setup}\label{sec:experiments-setup}
\subsubsection{User Profile} As shown in Figure~\ref{fig:pipeline} \textcolor[HTML]{B2B366}{(a)}, we generated a large number of user profiles with basic information, domain information, and context information using DeepSeek-R1, in order to support the simulation of diverse users. 

\subsubsection{Simulation Domains}
To evaluate the fast domain adaptability of \textbf{AdaptSim}, we conduct experiments in three representative domains: \textbf{Food}, \textbf{Beauty}, and \textbf{Shop}.

\begin{itemize}[label=\textbullet,leftmargin=*]
\item \textbf{Food Domain:} This scenario simulates an unconstrained food and restaurant recommendation system, where users can inquire about details such as taste, price, and reviews.
\item \textbf{Beauty Domain:} This scenario simulates an unconstrained cosmetics recommendation system, supporting user inquiries about details such as brand, efficacy, and usage experience.
\item \textbf{Shop Domain:} This scenario simulates an unconstrained electronics shopping recommendation system, where users can ask about products such as smartphones, laptops, and tablets, as well as details such as specifications, price, quality, and user reviews.
\end{itemize}

\subsubsection{User Simulator Baselines}
To evaluate our method, we select three advanced LLM-based user simulators as baselines, namely: 

\begin{itemize}[label=\textbullet,leftmargin=*]
\item \textbf{iEvaLM}~\cite{wang2023rethinking}: Generates user responses based on fixed templates with the target item as the user's profile.
\item \textbf{CSHI}~\cite{zhu2025llm}: Combines user profiles and action spaces to model user behavior.
\item \textbf{RecUserSim}~\cite{chen2025recusersim}: Extends CSHI with a larger action space and linguistic style modeling.
\end{itemize}

\subsubsection{Metrics for Evaluating Simulator}
We use the six-dimensional evaluation framework from RecUserSim~\cite{chen2025recusersim}. These six metrics were chosen to thoroughly evaluate the user simulator's ability to replicate real user behavior in interactions with a recommendation system. Naturalness ensures human-like dialogue, making it easier for the system to understand. Adaptability reflects the dynamic nature of user responses. Clarity ensures straightforward communication, while realism guarantees the simulator behaves like a real user. Relevance keeps the dialogue coherent and aligned with the context, and role-play ability ensures the simulator maintains its role as a user, not a recommender. 
\begin{itemize}[label=$\circ$,leftmargin=*]
  \item \textbf{Naturalness}: The extent to which the language generated by the user simulator resembles natural human language in terms of fluency, tone, and syntactic structure.
  \item \textbf{Adaptability}: The ability of a user simulator to dynamically adjust its requests or feedback in response to varying actions taken by the recommendation system, reflecting behavior that closely resembles that of a real user.
  \item \textbf{Clarity}: The degree to which the dialogues generated by a user simulator convey information clearly, directly, and effectively, ensuring ease of understanding.
  \item \textbf{Realism}: The extent to which a user simulator exhibits behavior that is believable and consistent with how a real human user would respond in a given interaction scenario.
  \item \textbf{Relevance}: The degree of coherence and logical consistency in the dialogue generated by the user simulator, maintaining semantic alignment and contextual continuity.
  \item \textbf{Role-play Ability}: The capability of the user simulator to consistently act as a user, engaging in behaviors such as seeking recommendations, responding to queries, and providing feedback.
\end{itemize}

\begin{figure}[t]
  \centering
  \includegraphics[width=1\linewidth]{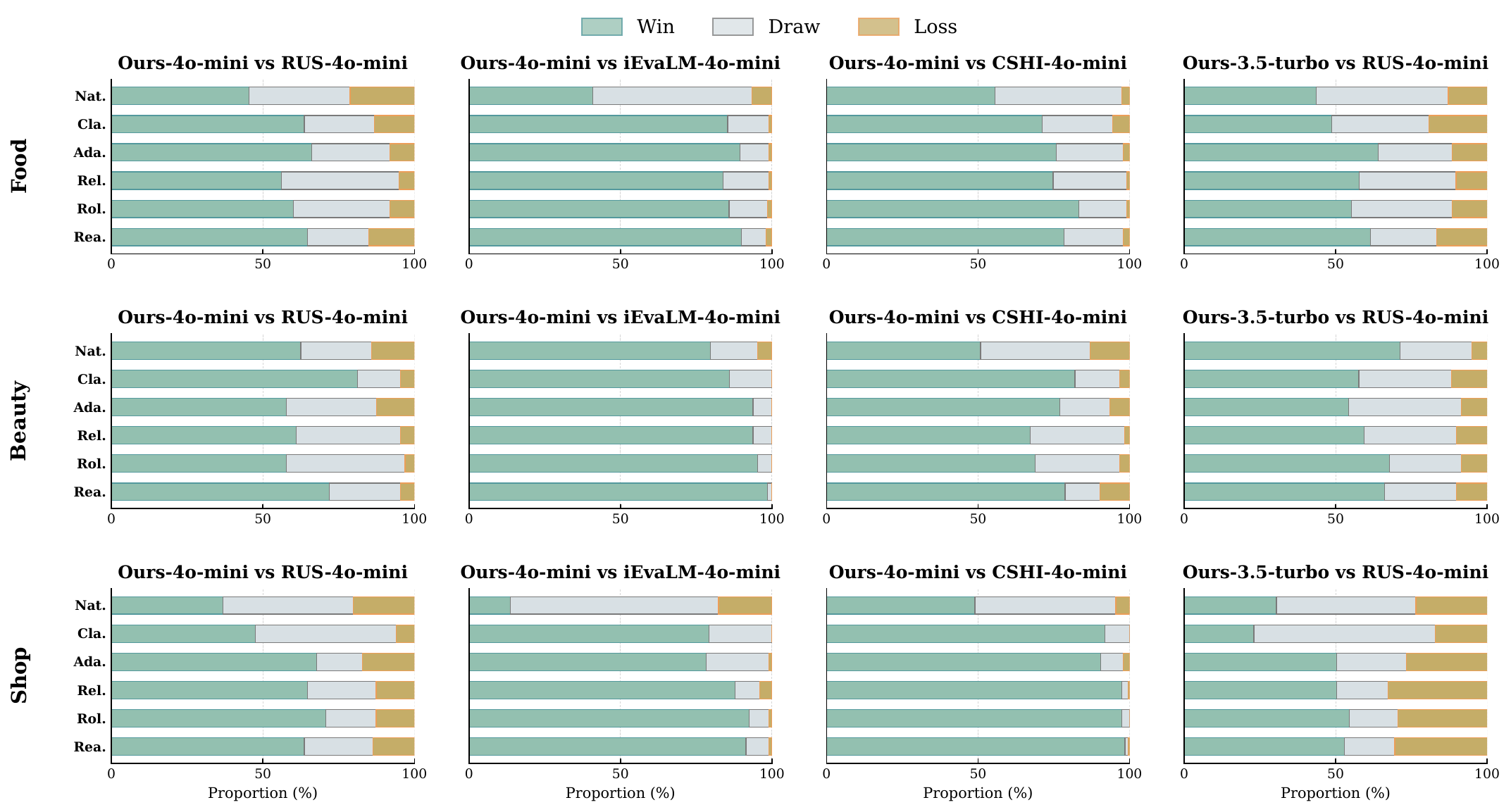}
\caption{Pairwise win--draw--loss comparison of AdaptSim against three baseline simulators across the Food, Beauty, and Shop domains. Each panel reports six dimensions: naturalness (Nat.), clarity (Cla.), adaptability (Ada.), relevance (Rel.), role-play ability (Rol.), and realism (Rea.). The four settings are AdaptSim with GPT-4o-mini versus RecUserSim, iEvaLM, and CSHI, and AdaptSim with GPT-3.5-turbo versus RecUserSim. Higher win proportions indicate better performance.}
  \label{fig:overall}
\end{figure}

\begin{figure}[t]
  \centering
  \includegraphics[width=1\linewidth]{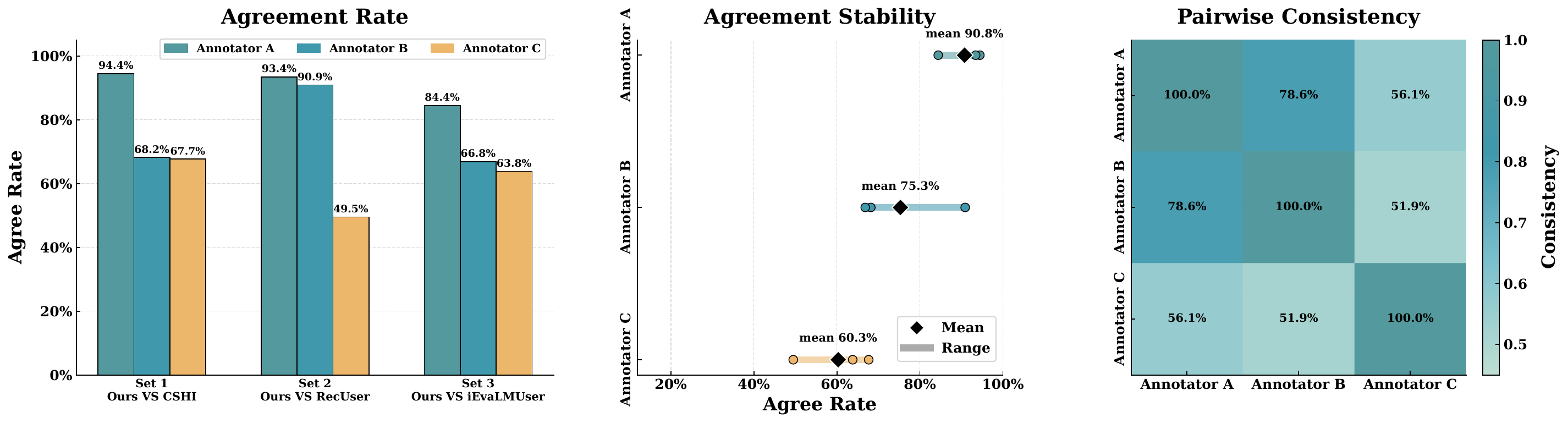}
\caption{Human validation of the LLM-as-a-judge protocol. Three senior annotators in the recommender systems domain reviewed the pairwise judgments and marked each case as agreement or disagreement. The left panel reports annotator-wise agreement rates across the three comparison sets, the middle panel summarizes their mean agreement rates and ranges, and the right panel reports pairwise inter-annotator consistency. The results show that all annotators agree with the LLM outputs in more than 60\% of cases on average, with moderate-to-high inter-annotator consistency, supporting the reliability of LLM-based evaluation.}
  \label{fig:agreement}
\end{figure}

\subsubsection{Language Style Details of Fine-Grained Control}
\begin{itemize}[label=$\circ$,leftmargin=*]
  \item \textbf{Formality}:  
  Value Range: (Formal, Informal)  
  Describes whether the generated language adheres to formal linguistic conventions or is more conversational and informal.

  \item \textbf{Sentiment}:  
  Value Range: (Positive, Neutral, Negative)  
  Describes the emotional tone conveyed by the language.

  \item \textbf{Conciseness}:  
  Value Range: (Concise, Moderate, Verbose)  
  Describes the length and information density of the user's language in expression.

  \item \textbf{Carelessness}:  
  Value Range: (Careless)  
Describes whether the language contains spelling mistakes, grammatical errors, incoherent expressions, and so on. 

\end{itemize}

\begin{figure}[t]
  \centering
  \includegraphics[width=1\linewidth]{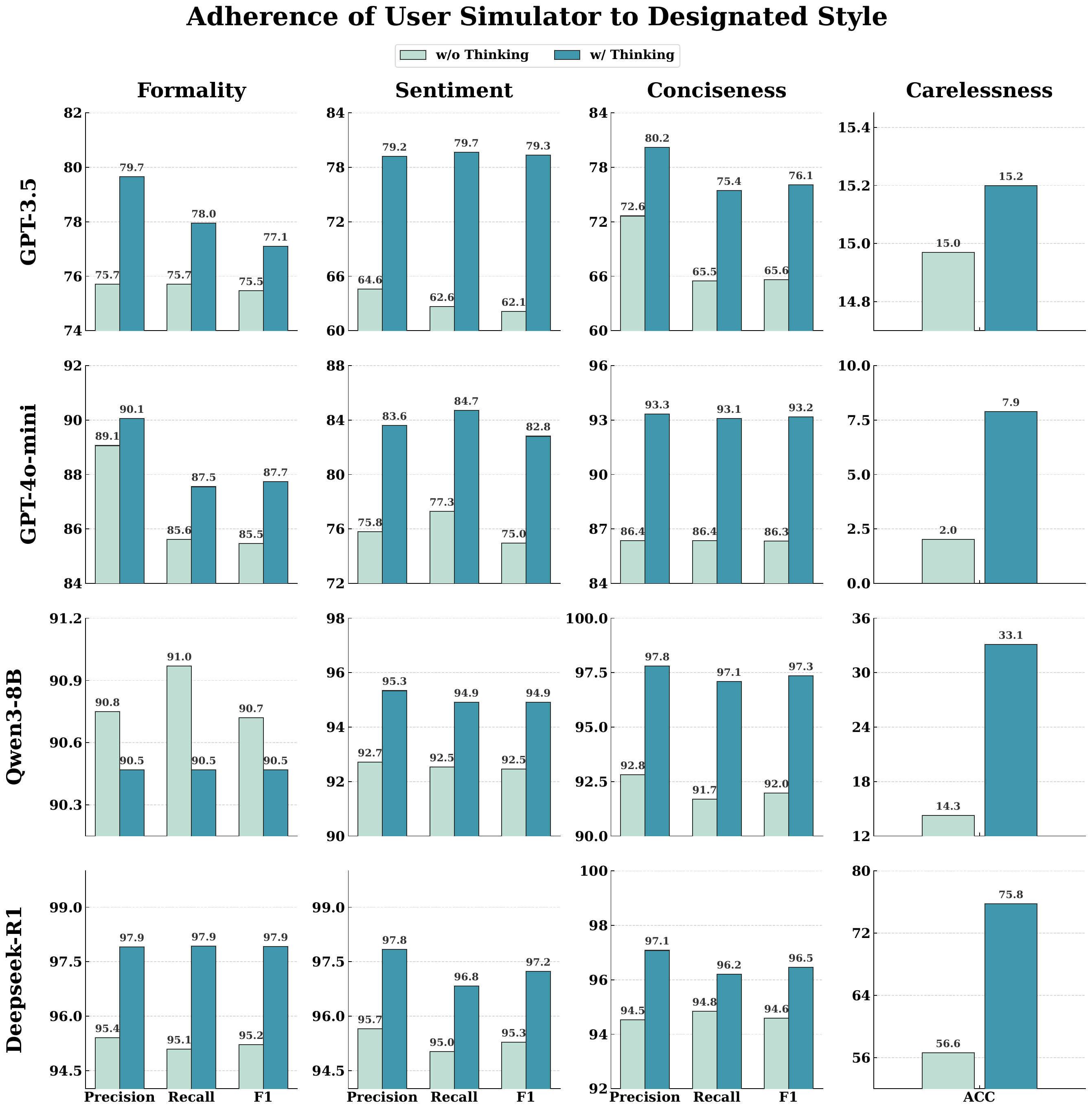}
    \caption{The adherence of the user simulator to the designated style across four LLM backbones and four style dimensions. Rows correspond to different backbones, while columns report formality, sentiment, conciseness, and carelessness. Entries marked (w) indicate that the model performs reasoning before generating a response, while (w/o) entries indicate direct response generation without intermediate reasoning. All values are percentages.}
  \label{fig:model_style}
  \vspace{-1em}
\end{figure}

\subsubsection{Metrics for Evaluating CRS}
The selection of these three metrics aims to assess the CRS's performance in language, actions, and recommendations. \textbf{Language Quality} ensures the dialogue is fluent, enhancing user experience. \textbf{Action Quality} evaluates how well the system interprets user intent and selects relevant responses. \textbf{Recommendation Quality} measures how well the system’s suggestions align with user preferences, reflecting the effectiveness of the recommendation mechanism.
\begin{itemize}[label=$\circ$,leftmargin=*]
  \item \textbf{Language Quality}: The extent to which the dialogue generated by the system demonstrates natural, fluent, and clear language use, resembling human-like communication.
  \item \textbf{Action Quality}: The system's ability to appropriately interpret user intent and select coherent, contextually relevant dialogue actions during the interaction process.
  \item \textbf{Recommendation Quality}: The degree to which the items suggested by the system align with the user's preferences and needs, reflecting the effectiveness of the recommendation mechanism.
\end{itemize}

\subsubsection{CRS Used for Evaluation}
For the food and beauty recommendation scenario, we evaluate two CRS frameworks~\cite{li2023large,chen2025recusersim}: 
\begin{itemize}[label=\textbullet,leftmargin=*]

\item \textbf{BaseCRS:} A single-prompt CRS that generates recommendations from dialogue history, serving as a simple baseline.
\item \textbf{AgentCRS:} A modular CRS with components for planning, memory (user preferences), and actions (ask, recommend, answer) guided by prompts.
\end{itemize}

\subsubsection{Implementation Details}
To evaluate the effectiveness of our proposed method, we incorporate a variety of LLMs as the backbone of user simulators, including Qwen3-8B~\cite{yang2025qwen3}, GPT-4.1-mini, DeepSeek-R1~\cite{guo2025deepseek}, and others. To ensure reproducibility, we set the temperature to 0. We use GPT-4o-mini to generate and optimize simulator prompts through meta-prompting. Since existing user simulators are often tailored to domains such as movies, we adapt and reconstruct their prompts to better align with our target domain.
We adopt a pairwise evaluation strategy using GPT-4o-as-a-judge~\cite{zheng2023judging} to compare AdaptSim against baseline methods. 

\begin{wrapfigure}{r}{0.58\linewidth}
  \vspace{-0.8em}
  \centering
  \includegraphics[width=1\linewidth]{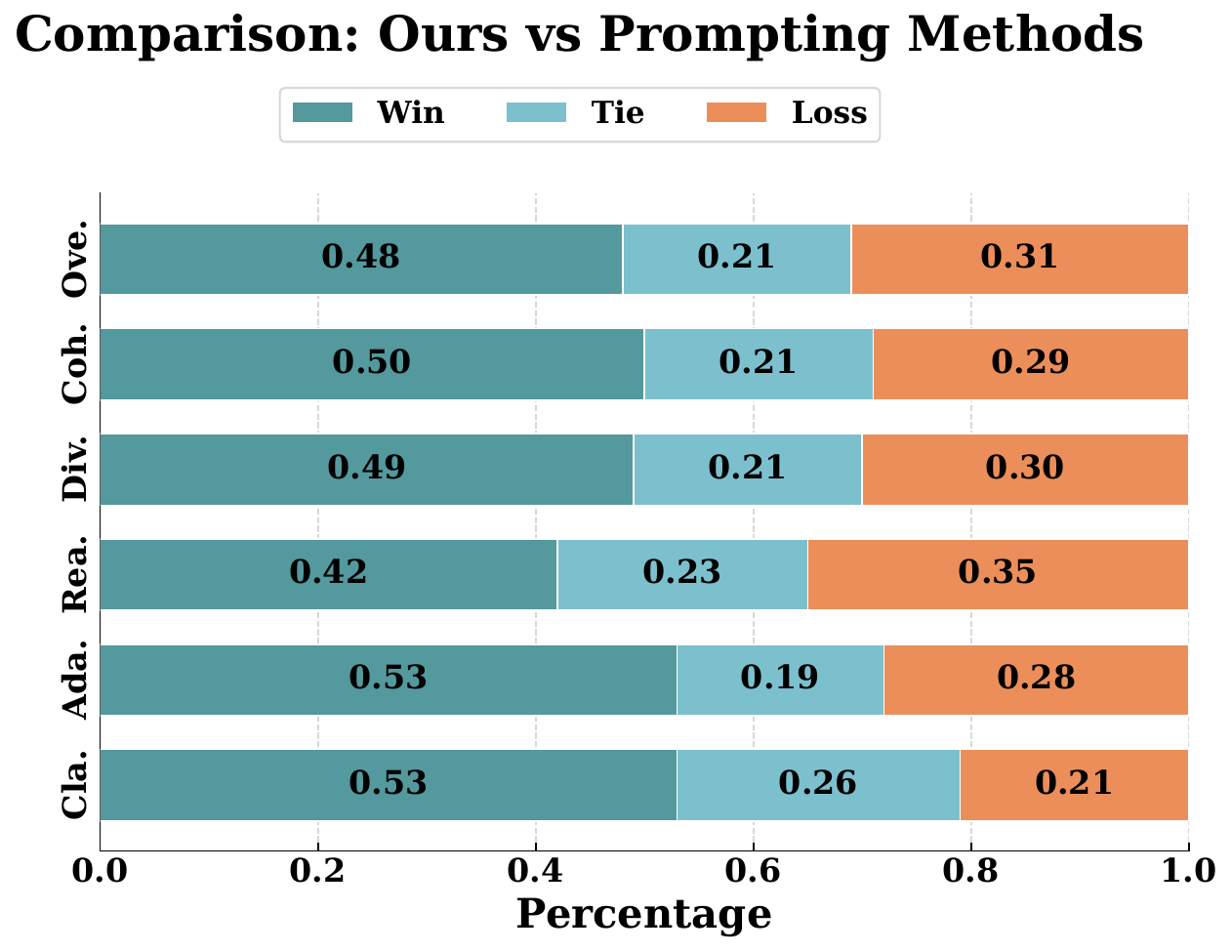}
  \caption{Comparison of AdaptSim vs. Prompting Methods in Chit-Chat Conversations. The metrics include overall performance, coherence, diversity, realism, adaptability, and clarity.}
  \label{fig:chit-chat}
\end{wrapfigure}

\subsection{Evaluation for User Simulator (RQ1)}\label{sec:user-simulator-eval}

\subsubsection{Evaluation on Simulation Quality (RQ1.1)}

Figure~\ref{fig:overall} compares AdaptSim with multiple baseline simulators across the Food, Beauty, and Shop domains using a pairwise win--draw--loss protocol. With the GPT-4o-mini backbone, AdaptSim achieves consistently higher win proportions across all six evaluation dimensions, and the green segments in the figure clearly reflect this advantage. We further evaluate AdaptSim with the weaker GPT-3.5-turbo backbone. Although performance declines as expected, AdaptSim remains competitive and still outperforms the baselines on most dimensions. The only clear exception is naturalness, where the results are closer to those of the baselines. This is unsurprising, as naturalness depends more directly on the underlying language generation ability of the backbone model than on the simulation framework itself. Overall, the results suggest that AdaptSim is effective, adaptable, and efficient, making it a strong choice for diverse simulation settings.

To assess the reliability of the LLM-as-a-judge protocol, we recruited three annotators with expertise in recommender systems to manually verify the judge outputs. For each pairwise comparison, they indicated whether they agreed or disagreed with the LLM judgment. As shown in Figure~\ref{fig:agreement}, the mean agreement rates of the three annotators are 90.8\%, 75.3\%, and 60.3\%, respectively, and their pairwise consistency is also moderate to high. These results suggest that the LLM-based evaluation is consistent with expert human judgment and is therefore reliable for our experiments.

\textbf{(Answer to RQ1.1)} AdaptSim outperforms all baselines in simulation quality across all three domains.

\subsubsection{Style Controllability Evaluation (RQ1.2)}

In this section, we assess a critical capability of AdaptSim: its fine-grained control over linguistic style. We evaluate its performance across four dimensions: formality, sentiment, and conciseness, which are measured by Precision, Recall, and F1-score, and carelessness, measured by accuracy. The results, presented in Figure~\ref{fig:model_style}, highlight the efficacy of our ``think-then-respond'' strategy. For conventional styles, AdaptSim demonstrates robust control across all tested LLM backbones. This is particularly evident with the DeepSeek-R1 model, which achieves exceptional F1-scores exceeding 95\% in formality, sentiment, and conciseness. 
Simulating unconventional styles like ``carelessness'' poses a greater challenge, primarily due to the inherent tendency of LLMs to produce coherent and polished text. Despite this intrinsic difficulty, the ``think-then-respond'' strategy significantly enhances performance, elevating the carelessness simulation accuracy to a notable 75.79\%. Collectively, these findings confirm that AdaptSim possesses strong and versatile style control capabilities, a crucial feature for generating diverse and realistic user behaviors for more nuanced conversational evaluations. 

\textbf{(Answer to RQ1.2)} AdaptSim provides effective fine-grained control over user linguistic styles.

\begin{figure}[t]
  \centering
  \includegraphics[width=1\linewidth]{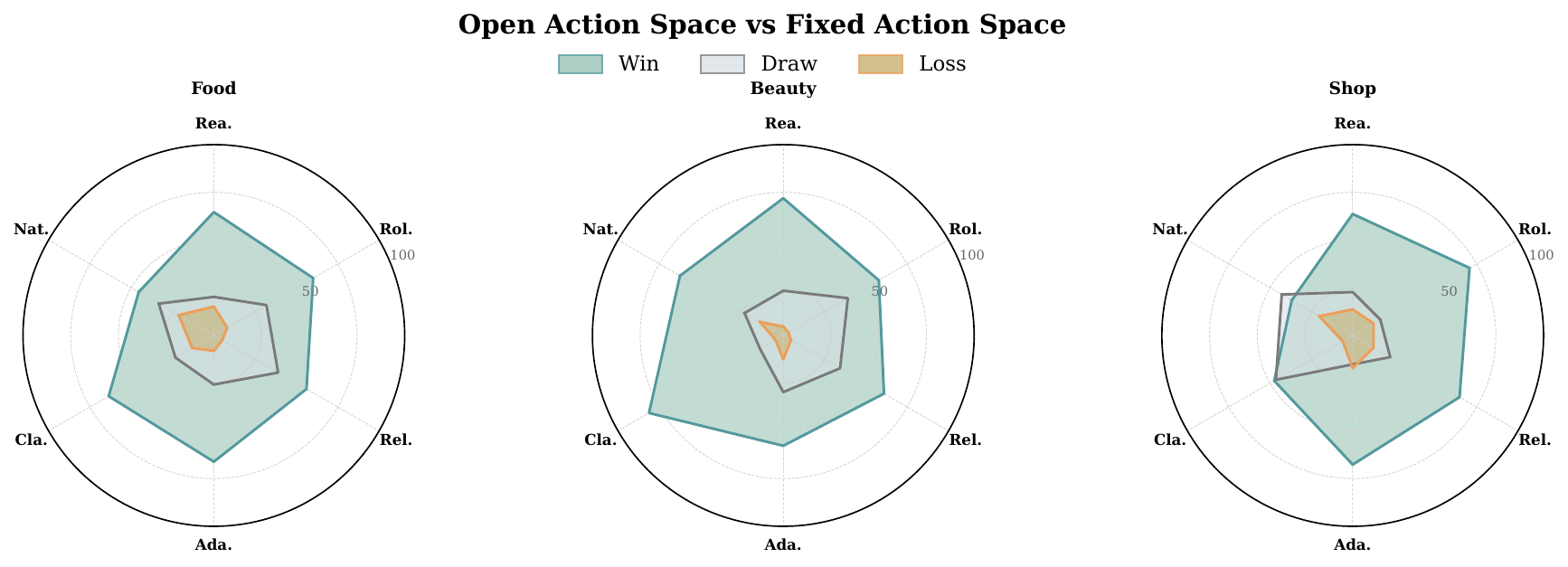}
\caption{Ablation analysis of open-ended action generation. The figure compares an open action space with a fixed action space across the Food, Beauty, and Shop domains using win--draw--loss proportions. Each radar chart reports six evaluation dimensions: naturalness (Nat.), clarity (Cla.), adaptability (Ada.), relevance (Rel.), role-play ability (Rol.), and realism (Rea.). Larger win regions indicate stronger performance of the open action space.}
  \label{fig:ablation}
  \vspace{-0.5em}
\end{figure}

\subsubsection{Using AdaptSim for Chit-Chat Conversations (RQ1.3)}

To rigorously test the extensibility and domain-agnostic nature of our framework, we adapted AdaptSim for the open-ended domain of chit-chat systems. This was achieved by systematically removing all domain-specific information from its prompts, creating a generalized version of the simulator. We then deployed this version to interact with personalized chatbots and benchmarked its performance against a strong baseline using pure prompting methods. The multi-dimensional evaluation, visualized in Figure~\ref{fig:chit-chat}, reveals a decisive advantage for our approach. AdaptSim not only achieves a higher overall performance but also consistently surpasses the prompting-based method across a spectrum of crucial conversational qualities, including coherence, diversity, realism, adaptability, and clarity. This is a significant finding, as it provides strong empirical evidence that the core mechanisms of AdaptSim are not narrowly tailored to task-oriented domains like CRS but are broadly applicable, showcasing its potential to enhance a wide array of advanced dialogue systems. 

\textbf{(Answer to RQ1.3)} AdaptSim generalizes well to open-ended chit-chat settings.


\begin{table}[t]
  \centering
\caption{Pairwise evaluation results of AdaptSim using different prompt sources: initial prompt (IP), human-crafted prompt (HP), and optimized prompt (OP). The ``Model'' refers to the backbone used by AdaptSim. The evaluation metrics include naturalness, clarity, relevance, roleplay, realism and adaptability}
\vspace{-0.8em}
  \label{tab:meta-prompt eff}
    \resizebox{\columnwidth}{!}{
  \begin{tabular}{ccccccccccc}
  \toprule
   & \textbf{Model}&\textbf{Result} & \textbf{Naturalness} & \textbf{Clarity}& \textbf{Relevance} & \textbf{Roleplay} & \textbf{Realism} & \textbf{Adaptability}\\
  \midrule
    AdaptSim (OP) & \multirow{3}{*}{3.5-turbo}& Win & 35.75 & 39.25 & 32.50 & 19.25 & 46.25 & 42.50 \\
  VS & & Tie& 45.75 & 29.50 & 49.75 & 66.25 & 26.00 & 27.00 \\
  AdaptSim (IP) & & Loss & 18.50 & 31.25 & 17.75 & 14.50 & 27.75 & 30.50 \\
  \midrule
    AdaptSim (OP)& \multirow{3}{*}{4o-mini}&Win& 31.50 & 40.50 & 30.75 & 17.00 & 46.75 & 43.50 \\
  VS & & Tie& 57.00 & 32.50 & 57.25 & 70.75 & 31.00 & 29.50 \\
  AdaptSim (IP)  &&Loss & 16.75 & 32.25 & 17.25 & 17.50 & 27.50 & 32.25 \\
\midrule
  AdaptSim (OP) &\multirow{3}{*}{4o-mini} & Win & 28.00 & 30.50 & 30.00 & 15.75 & 43.50 & 37.75 \\
  VS & & Tie & 54.50 & 28.25 & 54.50 & 68.75 & 28.50 & 28.50 \\
    AdaptSim (HP) &  & Loss & 17.50 & 41.25 & 15.50 & 15.50 & 28.00 & 33.75 \\
  \bottomrule
  \end{tabular}
  }
  \vspace{-1em}
\end{table}

\begin{figure}[t]
  \centering
  \includegraphics[width=1\linewidth]{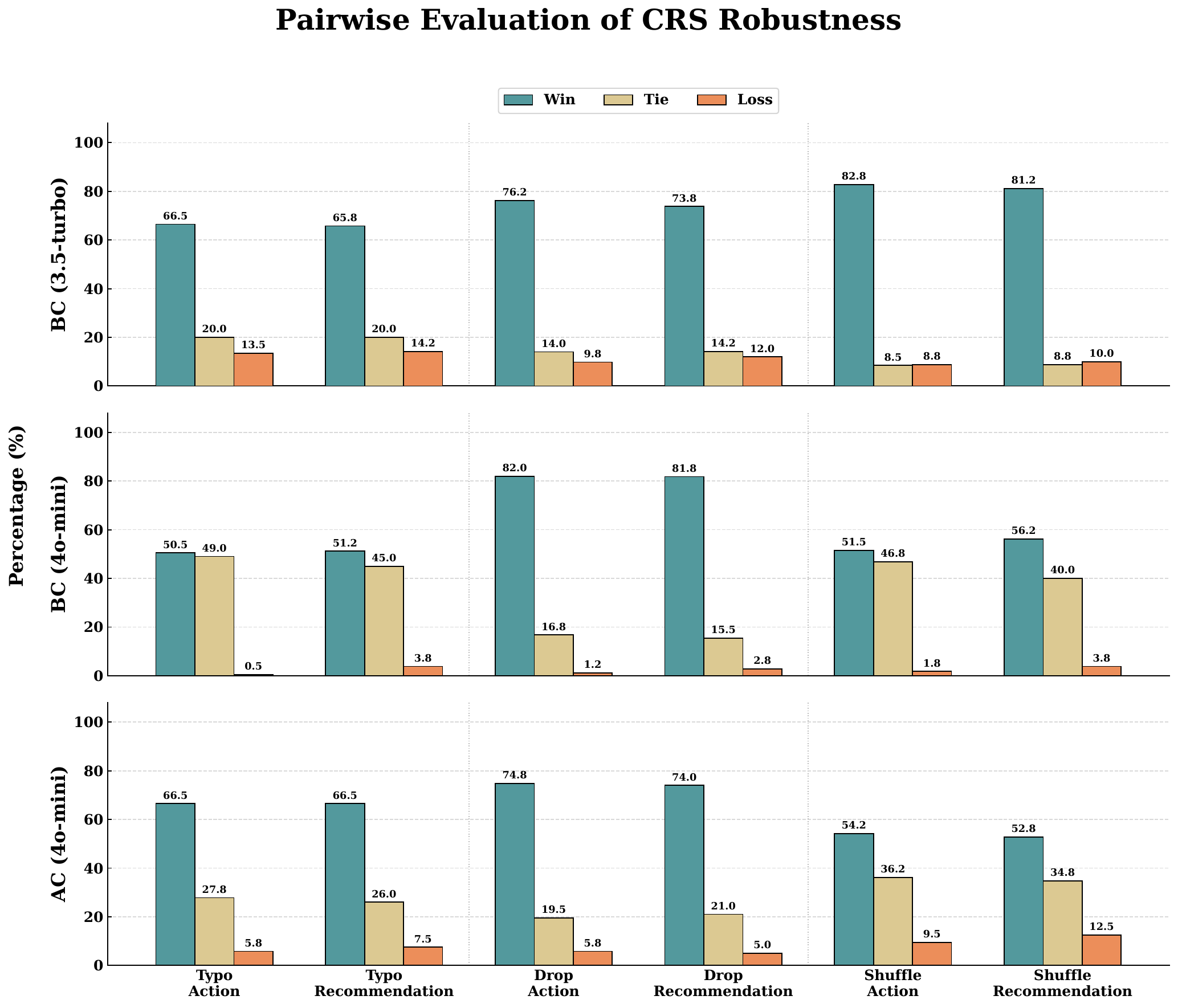}
\caption{Pairwise evaluation of CRS performance under normal and careless user conditions. Here, BC denotes BaseCRS; AC, AgentCRS; Rec, Recommendation; and 3.5-turbo and 4o-mini are shorthands for the base LLMs GPT-3.5-turbo and GPT-4o-mini.}
  \label{fig:crs_rob}
  \vspace{-1em}
\end{figure}

\subsubsection{Ablation Study (RQ1.4)}
In this section, we perform a series of ablation studies to examine the contribution of each component in \model, namely automatic prompt optimization (APO), Open-Ended Action Generation (OAG), and thinking-enhanced style generation (TSG). 
\begin{itemize}[leftmargin=*]
    \item \textbf{w/o APO}: This variant removes automatic prompt optimization and directly uses the initial prompt, $\mathrm{Prompt}_{0}$, for dialogue generation. To evaluate APO, we compare this variant with full AdaptSim using the optimized prompt and include a human-crafted prompt as an additional reference. As shown in Table~\ref{tab:meta-prompt eff}, the optimized prompt consistently outperforms the initial prompt under both GPT-3.5-turbo and GPT-4o-mini backbones. It also often matches or exceeds the human-crafted prompt, indicating that APO effectively improves prompt quality and reduces dependence on manual prompt engineering.
    \item \textbf{w/o OAG}: This variant removes open-ended action generation and forces the simulator to rely on a fixed action space. As shown in Figure~\ref{fig:ablation}, the open action space achieves larger win proportions than the fixed action space across the Food, Beauty, and Shop domains, covering most evaluation dimensions. This suggests that OAG enables more flexible and context-aware user actions, improving adaptability, relevance, and overall simulation quality.
    \item \textbf{w/o TSG}: This variant removes thinking-enhanced style generation and directly produces the response without an explicit intermediate reasoning process. As shown in Figure~\ref{fig:model_style}, removing this component lowers style adherence across all tested styles and backbones. In contrast, the full model with thinking consistently achieves higher Precision, Recall, and F1 for formality, sentiment, and conciseness, while also improving accuracy on the carelessness setting. This indicates that intermediate reasoning plays an important role in preserving the designated style during generation.
\end{itemize}

\textbf{(Answer to RQ1.4)} The consistent performance degradation across ablation variants confirms that automatic prompt optimization, open-ended action generation, and thinking-enhanced style generation are all important for the optimal performance of \model.

\subsubsection{Case Study (RQ1.5)}
We further conduct case studies to examine the practical behavior of existing user simulators, including RecUserSim, iEvaLM, and CSHI. As shown in Figures~\ref{tab:case1_recusersim}--\ref{tab:case_shot_style_ours}, these simulators commonly suffer from role reversal, where the simulated user mistakenly acts as the recommender during the interaction. This problem is closely related to their design limitations. For example, iEvaLM lacks explicit action guidance, which makes its generated behavior less goal-directed, while RecUserSim does not sufficiently exploit dialogue context during refinement and can therefore lose track of the user role.

We also observe frequent repetition of identical or highly similar questions, indicating weak context tracking and limited dialogue coherence. This issue is particularly likely when a simulator relies on a fixed or narrow action space, as in CSHI and RecUserSim, which can repeatedly select similar actions across consecutive turns. In contrast, AdaptSim maintains the predefined user persona and linguistic style throughout the interaction, while using open-ended actions to better align with the dialogue context. As a result, it reduces role confusion and repetitive behavior, leading to more coherent and realistic user simulations.

\textbf{(Answer to RQ1.5)} AdaptSim mitigates common simulator failures such as role reversal and repetitive questioning in case studies.

\subsection{Evaluation for CRS (RQ2)}\label{sec:crs-eval}
\subsubsection{Evaluation of CRS Basics (RQ2.1)}

\begin{table*}[t]
    \centering
    \caption{Pairwise evaluation results of AdaptSim in assessing CRS models built with different base LLMs. AS, BC, and AC denote AdaptSim, BaseCRS, and AgentCRS, respectively, while 4.1-mini and 4o-mini are shorthands for GPT-4.1-mini and GPT-4o-mini.}
    \label{tab:crs_evaluation}
    \resizebox{\textwidth}{!}{
    \begin{tabular}{ccccccccccc}
    \toprule
    \multirow{2}{*}{} & \multirow{2}{*}{} & \multicolumn{3}{c}{ASim (4.1-mini)} & \multicolumn{3}{c}{ASim (4o-mini)} & \multicolumn{3}{c}{ASim (Qwen3-8B)}\\
    \cmidrule(r){3-5} \cmidrule(r){6-8} \cmidrule(r){9-11}
     & & Win & Tie & Loss & Win & Tie & Loss & Win & Tie & Loss  \\
    \midrule
    BaseCRS (GPT-4o-mini)& Action& 66.5 & 27.2 & 6.2 & 59.8 & 29.6 & 10.6& 64.8 & 24.8 & 10.5 \\
    VS & Language& 60.2 & 38.5 & 1.2& 59.8 & 35.6 & 4.7& 67.5 & 27.2 & 5.2 \\
    BaseCRS (3.5-turbo)& Rec & 65.0 & 27.5 & 7.5& 57.0 & 33.0 & 10.0& 63.8 & 28.2 & 8.0 \\
    \midrule
    BaseCRS (GPT-4o-mini) & Action& 40.5 & 47.5 & 12.0& 52.5 & 37.5 & 10.0& 61.5 & 28.0 & 10.5 \\
    VS & Language& 20.8 & 74.0 & 5.2& 33.0 & 60.0 & 7.0& 33.2 & 60.0 & 6.8 \\
    AgentCRS (GPT-4o-mini)& Rec& 43.5 & 38.2 & 18.2 & 48.5 & 35.5 & 16.0& 56.8 & 27.0 & 16.2\\
    \midrule
    AgentCRS (GPT-4o-mini)& Action&59.0 & 22.8 & 18.2& 52.5 & 22.2 & 25.2& 50.8 & 23.2 & 26.0\\
    VS& Language& 39.2 & 49.8 & 11.0& 36.2 & 50.2 & 13.5 & 41.0 & 42.8 & 16.2 \\
    AgentCRS (GPT-3.5-turbo)& Rec& 57.5 & 23.5 & 19.0& 52.2 & 26.0 & 21.8& 51.5 & 22.8 & 25.8
     \\
    \bottomrule
    \end{tabular}
    }
    \vspace{-1em}
    \end{table*}

\begin{figure}[t]
  \centering
  \includegraphics[width=1\linewidth]{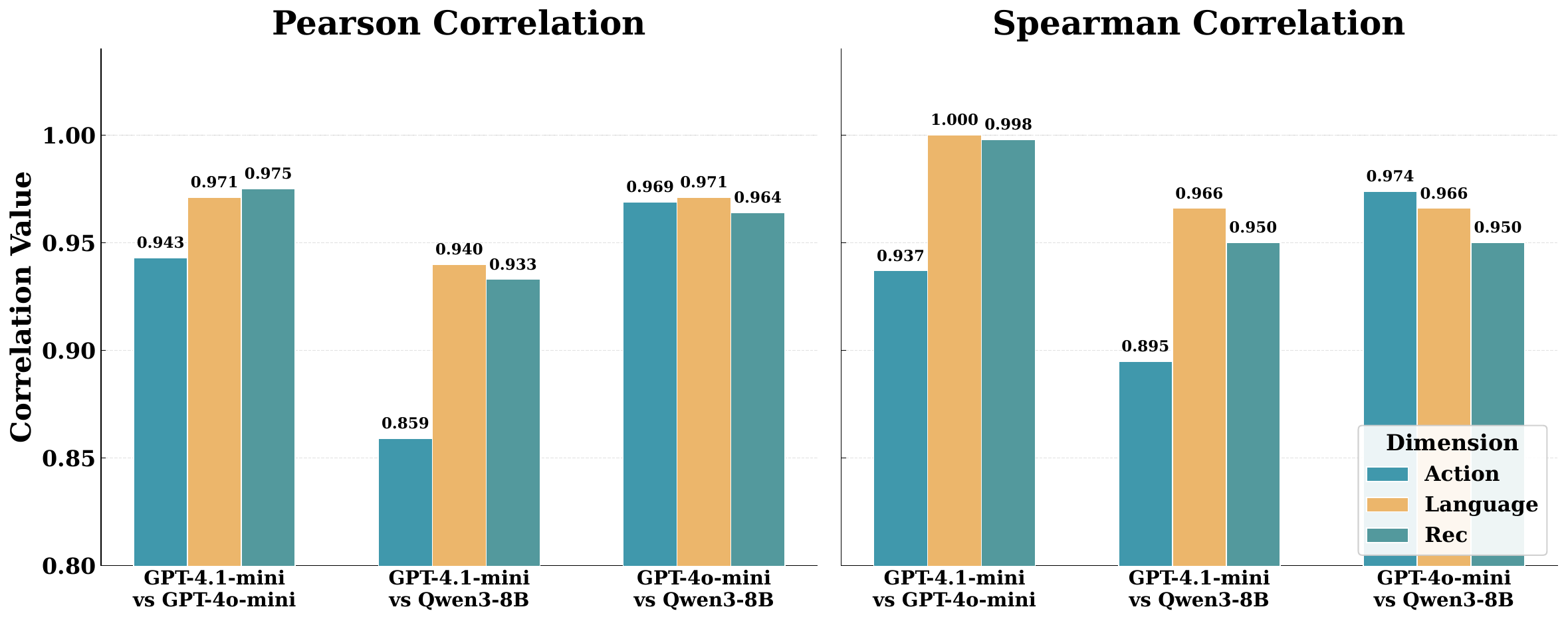}
\caption{Pearson and Spearman correlation of AdaptSim evaluation scores across backbone LLM pairs and task dimensions (Action, Language, Recommendation). All reported correlations are statistically significant ($p < 0.05$), indicating that the proposed evaluation protocol produces stable and consistent rankings regardless of the simulator backbone.}
  \label{fig:correlation}
  \vspace{-1em}
\end{figure}

Following the protocol in Section~\ref{sec:BFS_Bas}, we use \model\ to conduct turn-level pairwise comparisons between CRSs under matched dialogue contexts. We evaluate two representative architectures: BaseCRS, which directly generates responses from the dialogue history, and AgentCRS, which decomposes the interaction into planning, memory, and action modules. The comparisons cover three dimensions: action quality, language quality, and recommendation quality.
Table~\ref{tab:crs_evaluation} shows two consistent patterns. First, the backbone LLM has a clear effect on CRS quality. For both BaseCRS and AgentCRS, versions built on GPT-4o-mini obtain higher win rates than their GPT-3.5-turbo counterparts across action, language, and recommendation dimensions. This indicates that stronger language models improve not only surface-level response fluency, but also downstream decision making in conversational recommendation.
Second, architecture influences different capabilities in different ways. BaseCRS generally outperforms AgentCRS in action and recommendation quality, while language quality is often closer and contains many ties. One possible explanation is that the modular design of AgentCRS provides explicit structure, but its predefined action space can also restrict how flexibly the system reacts to diverse user intents. In contrast, BaseCRS is less structured but can adapt its response more freely to the current dialogue context.
To examine whether these conclusions depend on the simulator backbone, we further compare the rankings produced by \model\ with different backbone LLMs. As shown in Figure~\ref{fig:correlation}, the rankings are highly consistent across backbones, with Pearson correlations ranging from $0.859$ to $0.975$ and Spearman correlations ranging from $0.895$ to $1.000$. These results suggest that the proposed evaluation protocol provides stable comparative signals for CRS evaluation.

\textbf{(Answer to RQ2.1)} \model\ enables reliable evaluation of CRS core capabilities: it identifies differences caused by backbone models and CRS architectures, and its rankings remain consistent across simulator backbones.

\subsubsection{Evaluation of CRS Robustness (RQ2.2)}
We next evaluate whether current CRSs remain reliable when users produce imperfect inputs. Following Section~\ref{sec:BFS_Rob}, we use the fine-grained style control of \model\ to generate careless user utterances with typos, word shuffling, and word dropping. This setting simulates common interaction noise caused by informal typing, incomplete expressions, or low-effort user behavior.
We compare each CRS under clean and perturbed user inputs while preserving the same dialogue context for each turn-level comparison. As shown in Figure~\ref{fig:crs_rob}, noisy inputs consistently reduce CRS performance. The degradation appears across different CRS architectures and backbone LLMs, suggesting that robustness is not solved simply by using a stronger generator or a more modular CRS pipeline. Instead, existing systems remain vulnerable when user intent is expressed in a less polished form.
These findings demonstrate an important use case of \model: by controlling user language at a fine-grained level, it can expose failure modes that are difficult to observe with clean or template-based simulations. Robust CRS evaluation should therefore include not only standard recommendation success under ideal inputs, but also stress tests under realistic linguistic variation and noise.

\textbf{(Answer to RQ2.2)} \model\ enables controlled robustness evaluation for CRSs, revealing that current systems remain sensitive to noisy user inputs and require stronger fault tolerance.

\section{CONCLUSION}
\label{sec:conclusion}
In this paper, we studied LLM-based user simulation for conversational recommender systems, focusing on the limitations of existing simulators in domain adaptability, user modeling, and evaluation validity. We proposed \model, an adaptive user simulator that combines automatic prompt optimization, open-ended action generation, and thinking-enhanced style generation. These designs reduce the dependence on manually crafted prompts and predefined action spaces, while enabling more flexible behavior modeling and fine-grained control over user linguistic styles.

We further introduced a BFS-based turn-level pairwise evaluation framework for assessing the basic capabilities and robustness of CRSs under matched dialogue contexts. Extensive experiments across multiple domains and LLM backbones show that \model produces more realistic, adaptable, and controllable user simulations than existing baselines. The results also demonstrate that \model can support reliable CRS evaluation and reveal important robustness weaknesses of current CRS models under noisy user inputs.

\section{LIMITATIONS AND FUTURE WORK}
Although AdaptSim demonstrates strong performance across domains and backbones, several limitations tied to its specific design choices warrant further investigation.

First, the automatic prompt optimization (APO) process relies on an LLM both to detect behavioral issues in generated dialogues and to refine the prompt accordingly. This creates a circular dependency: the quality of the optimized prompt is bounded by the meta-evaluation LLM's ability to identify subtle simulation failures, such as underspecified intent or inconsistent guidance. When the backbone LLM is weak, both the error detection and the refinement steps may degrade simultaneously, making it difficult to diagnose whether poor simulation quality stems from the prompt or the model itself.

Second, the open-ended action generator selects the most probable candidate action, which favors high-frequency, prototypical user behaviors. This greedy selection may suppress rare but realistic actions---such as abrupt topic shifts or contradictory preference updates---that are important for stress-testing CRS robustness. Incorporating diversity-aware sampling or explicit coverage objectives could address this.

Third, the thinking-enhanced style controller enforces stylistic consistency through a prompt-driven thinking step, without modifying the model's internal representations. In very long dialogues, the style signal may still drift as the thinking step competes with an increasingly long context. A more robust solution might involve fine-tuning or retrieval-augmented style anchoring.

Finally, the BFS-based evaluation expands dialogue states breadth-first, which provides context-consistent pairwise comparisons but grows exponentially with branching factor and depth. In practice, we cap the depth at $R$ and limit branching, which may miss important late-turn failure modes. Future work could explore beam-search-style pruning or learned branch selection to improve coverage without proportional cost.


\clearpage
\appendix
      
          
          
          

          
          

\section{Additional Prompt Templates and Case Studies}\label{app:appendix}
This appendix provides the prompt templates used in AdaptSim and representative qualitative case studies. The prompt templates cover profile construction, action generation, response generation, issue detection, prompt refinement, and style control. The case studies illustrate typical failure modes of baseline user simulators, including role reversal and repetitive questioning, as well as representative examples of AdaptSim under different linguistic styles. These materials complement the quantitative results in the main text and provide additional detail for reproducibility and qualitative inspection.

\subsection{Prompt Templates}

\begin{itemize}[leftmargin=*]
  \item Profile prompt generation for constructing natural-language user profiles from structured fields.
  \item Initial action-generation prompt for open-ended user intent modeling.
  \item Response-generation prompt for turning intended actions into user utterances.
  \item Issue-discovery prompt for analyzing dialogue failures in user simulator outputs.
  \item Prompt-refinement prompt for improving action-generation quality based on detected issues.
  \item Final intent-action generation prompt with probability-based outputs and reasoning traces.
\end{itemize}

\PromptTitle{Meta prompt to generate profile prompts.}{tab:prompt1}
\begin{PromptBlock}
You are a Prompt Engineer. Your task is to construct a prompt template that generates a natural language user profile for a user simulator, given a domain and a set of structured fields.

Goal: Automatically express a user's preferences, dislikes, contextual signals, and constraints in fluent natural language.

Below is an example:

Input:
{
  "domain_sys": "Food Recommendation System",
  "domain_specific": [
    "locationNot", "tasteNot", "shopNot", "constraint",
    "time", "taste", "location", "listType", "scenes"
  ],
  "context": [
    "price", "rating", "privateRoom", "radius", 
    "sourceLocation", "targetLocation"
  ]
}

Output:
```jinja2
Domain Specific Language Info (Omitted due to space limitations)
```

Now, based on the following input fields, generate a Profile Prompt for a user simulator.

Input:
{Your profile fields (input)}

Output:
```jinja2
<Your Output>
```
\end{PromptBlock}

\vspace{+1em}
\PromptTitle{Initial Prompt Template Design for Action Generation in All Domains.}{tab:prompt2}
\begin{PromptBlock}
You are a highly realistic user simulator whose goal is to obtain satisfactory responses by interacting with {domain_sys}.

Here is some information about you:
{Basic Info}
{Domain Specific Info}
{Context Info}

Your historical conversation records with {domain_sys}, along with the actions you took in each turn, are as follows:
{Dialogue_Action_history}

As the user, please generate a set of appropriate actions you will take.
You should follow these rules:
1. No repetition: do not generate actions already present or semantically similar in the history.
2. If the system asks you for information, you should respond immediately with the corresponding action.

Please combine your personal information, history, and relevant actions to generate up to 5 actions with their probabilities, summing to 100%.
Your output must be strictly controlled to 3 lines:
Line 1: Reflect on whether you have taken any action that {domain_sys} cannot reply to, and whether you have answered all necessary questions from {domain_sys}.
Line 2: List all the actions you will take in the format: action1 : prob; action2 : prob; ... (avoid repeated or similar actions).
Line 3: Directly state the reason for choosing each action, in the same order as actions.

Your number of reasons must exactly equal the number of actions.

Output the results strictly in the following JSON format:
{
  "Thought": "<Did you take any actions that {domain_sys} cannot respond to? Did you answer all necessary questions from {domain_sys}?>",
  "Actions": "[action1:prob; ..., action5:prob]",
  "Reasons": "[action reason1; ...; action reason5]"
}
\end{PromptBlock}

\PromptTitle{Prompt design for response generation.}{tab:prompt3}
\begin{PromptBlock}
You are a user simulator. Your current task is to play the role of a real user and interact with the {domain_sys}. You now need to generate a response to {domain_sys}.

When generating this reply, please follow these instructions carefully:
1. Users typically do not repeatedly mention the same preferences unless the system continues to make incorrect recommendations. For example, if the user has already stated their food preference, they generally will not mention it again in later conversations.
2. In your reply, you must complete all the actions you plan to take. You cannot execute only part of them. If your intended action is a single one, your reply only needs to complete that one action; if it is a compound action, you need to complete all the actions in your reply.
3. Avoid replies that are too similar or identical to previous responses. The reasons are: if you have already asked a similar question and the system responded correctly, you have received the necessary information; if the system responded incorrectly, repeating the question will not yield a better answer.
4. If the recommendation system asks you for some information, but that action is not listed, you should still provide the information, because it is necessary.
5. Special case: if the action to be taken is ending the conversation, your reply must include farewell words such as "Goodbye" or "Bye", and the conversation should end. Do not ask the recommendation system any further questions.

Here is some information about you:
{Basic Info}
{Domain Specific Info}
{Context Info}

Your historical conversation records with {domain_sys}, along with the actions you took in each turn, are as follows:
{Dialogue_Action_history}

Your intended actions are: [{action_names}]
Your reasons for choosing these actions are: [{inner_voice}]

Now, please play the role of a user.
If this is the first round of conversation, you must initiate the dialogue. In this case, you do not need to respond to the recommendation system.
As a user, based on your information and intended actions, and strictly adhering to your language style, generate a response to the recommendation system.
Please follow this Language Style:
{Language Style}

Your output must be exactly 4 lines, structured as follows:
Line 1: Reflect on whether you followed your language style.
Line 2: Reflect on whether you illegally asked questions if the action is to end the conversation.
Line 3: Output a single line of user response to the system.
Line 4: Explain why you responded this way.

Output the results strictly in the following JSON format:
{
  "Thought1": "<What is your language style? Did you follow it?>",
  "Thought2": "<If your current action is to end the conversation, did you illegally ask {domain_sys} any questions?>",
  "Response": "Your reply to the system.",
  "Reason": "Why did you respond this way?"
}
\end{PromptBlock}

\PromptTitle{Meta prompt for identifying issues in user simulator dialogues.}{tab:prompt4}
\begin{PromptBlock}
You are a problem discovery expert, specializing in analyzing the quality of dialogues between a user simulator and a CRS (Conversational Recommender System).

Your task is: given five dialogues between the user simulator and the CRS, identify the potential issues in the user simulator's behavior during the conversation.

Please conduct an in-depth analysis across the following dimensions:
1. Behavioral Consistency: whether the user's responses deviate from reasonable behavioral patterns or role settings.
2. Logical Coherence: whether there are logical contradictions, semantic jumps, or repetitive content.
3. Diversity and Naturalness: whether the user expressions lack realism, diversity, or appear overly templated.
4. Language Quality: whether the language is mechanical, rigid, or inconsistent with everyday human communication habits.
5. Context Understanding Ability: whether the user shows unreasonable neglect or misunderstanding of prior context.

The five dialogues are as follows:
{% for dialogue in dialogues %}
Dialogue {{ loop.index }}:
{{ dialogue }}
{% endfor %}

Do not base your problem descriptions on the original dialogues directly; aim to identify high-level issues.

Please output the results in the following format:
{
  "Analysis": "A brief summary of the overall 
  issues and trends observed across the five dialogues",
  "Issues": [
    "Issue: Issue description",
    "Issue: Issue description",
    ...
  ]
}
\end{PromptBlock}

\vspace{+1em}
\PromptTitle{Meta prompt for optimizing user simulator's action generation.}{tab:prompt5}
\begin{PromptBlock}
You are a senior Prompt Optimization Expert, specializing in refining prompts for user simulators to enhance the realism and interaction quality of generated user intentions.

Background: The current user simulator uses a Prompt to generate the user's intention, which is followed by another Prompt to generate the user's actual response. This task focuses solely on improving the first Prompt — the one responsible for generating user intentions. Your goal is to enhance the realism, naturalness, and coherence of these intentions, while avoiding common issues such as illogical flow, unnatural tone, or rigid/templated expressions.

Important: The Prompt you optimize is only responsible for producing the user's initial intention, not their actual reply.

Please complete the following tasks:
1. Carefully review the Initial Prompt and the accompanying Identified Issues from previously generated dialogues.
2. Analyze the probable causes behind the issues — which parts of the Prompt are vague, overly rigid, misaligned with the goal, or lack appropriate constraints.
3. Rewrite and refine the Prompt so that it more effectively guides the simulator to produce user intentions that are realistic, consistent with human behavior, naturally phrased, and coherent with the dialogue context.
4. Ensure that the generated user intention represents a clear and reasonable request for recommendation.

You will be provided with:
Identified Issues with the User Simulator:
{{ problems }}

Initial Prompt:
{{ initial_prompt }}

Before writing the final output, reflect on what the initial prompt is trying to achieve.

Please provide your output in the following format:
Reflection: Your analysis and insights here

Optimized Prompt:
<jinja2>
Refined Prompt content
</jinja2>
\end{PromptBlock}

\PromptTitle{Prompt design for realistic user intent action generation.}{tab:prompt6}
\begin{PromptBlock}
You are a highly realistic user simulator whose goal is to obtain satisfactory responses from {{ domain_sys }} by engaging in dialogue with it using the thinking and expression patterns of real users. Your behavior should reflect the diversity, emotions, feedback, and personality of real users, and avoid mechanical, templated, or repetitive expressions.

Here is some information about you:
{Basic Info}
{Domain Specific Info}
{Context Info}

Your historical conversation records with {domain_sys}, along with the actions you took in each turn, are as follows:
{Dialogue_Action_history}

You and {{domain_sys}} have the following historical conversation and user actions per round:  
{{chat_action_his}}

Please generate up to 5 "user intent actions" and their probabilities (summing to 100%) from the perspective of a real user, based on your personal information, the conversation history, and the system's responses. Your behavior should meet the following requirements:
1. In the first round, you must proactively request a recommendation, expressing your needs or confusion, combined with your lifestyle, preferences, budget, owned products, etc., using natural, colloquial, and life-like expressions, avoiding templated or formal language.
2. In subsequent rounds, prioritize feedback, evaluation, follow-up questions, or supplementary information based on the system's response, reflecting real user emotions, doubts, approval, rejection, indecisiveness, price sensitivity, brand preference, etc., avoiding mechanical questioning.
3. Behavior should be logically progressive, avoiding repetition of past actions or semantically similar ones, and demonstrating absorption, filtering, and personal refinement of previous information.
4. You may appropriately display hesitation, vague expressions, forgotten information, misunderstandings, or colloquial phrases typical of real users to increase naturalness and challenge.
5. At the end, avoid templated goodbyes; you may express gratitude, hesitation, additional questions, or genuine feedback on the recommendations based on actual experience.
6. Each action must have a clear and specific reason, which should be personalized and contextualized based on your background, the conversation history, and real user psychology.

Your output must strictly be in 3 lines:
Line 1: Reflect on whether you have taken actions that {{domain_sys}} cannot respond to, and whether you have answered all necessary questions from {{domain_sys}}.
Line 2: List all actions and their probabilities, formatted as [action1:prob; ..., action5:prob], with actions being diverse, natural, and non-repetitive.
Line 3: Provide the specific reasons for each action, matching the number of actions, with reasons being specific, personalized, and contextually relevant.

Output the results strictly in the following JSON format:
{
  "Thought": "<Did you take any actions that {domain_sys} cannot respond to? Did you answer all necessary questions from {domain_sys}?>",
  "Actions": "[action1:prob; ..., action5:prob]",
  "Reasons": "[action reason1; ...; action reason5]"
}
\end{PromptBlock}


\subsection{Representative Case Studies}
The following examples highlight common failure modes of baseline user simulators and illustrate how AdaptSim maintains more coherent role behavior and more controlled language generation across dialogue settings.

\begin{itemize}[leftmargin=*]
  \item RecUserSim bad cases showing role reversal under the GPT-4o backbone.
  \item iEvaLM bad cases showing repeated questioning and weak strategic guidance.
  \item CSHI bad cases showing repeated questioning under a limited action space.
  \item AdaptSim case studies for formal, informal, long, and short linguistic styles.
\end{itemize}

\begin{figure*}[h]
  \centering
  \begin{tcolorbox}[
      width=\textwidth,           
      enhanced,                   
       colframe=blue!50!black,     
      colback=blue!10!white,       
      colbacktitle=blue!60!black, 
      coltitle=white,             
      title=\large \textbf{Case Study: RecUserSim (GPT-4o) BadCase 1},   
      fonttitle=\bfseries,
      arc=4mm,                    
      boxsep=5pt,                 
      top=5pt, bottom=5pt,        
    fontupper=\small,
  ]
\textbf{User:} Can you recommend some good places to eat near me? Thank you! \\ [2ex]
    \textbf{Assistant:} Of course! To give you more personalized recommendations, could you tell me which city or area you're in? Also, do you have a preference for a certain type of cuisine, like Chinese, Western, Japanese, or something else? Any budget range or special requirements? \\[2ex]
    \textbf{User:} I recommend you try the nearby fast food restaurants — they usually offer a variety of quick meals that should suit your needs.\textcolor{blue}{\textbf{(Bad role play ability-- role reversal)}} \\ [2ex]
    \textbf{Assistant:} Recommending fast food is a great idea! If you'd like to try different kinds of fast food, here are some popular chain restaurants you can consider:
    
    - \textbf{McDonald's} – Classic burgers, fries, and a wide breakfast menu.
    
    - \textbf{KFC} – Famous for fried chicken and various side dishes, suitable for different tastes.
        
    - \textbf{Subway} – Custom sandwiches and healthy salads, perfect for a quick and healthy meal.
        
    - \textbf{Burger King} – Known for flame-grilled burgers and also offers other snack options.
        
    - \textbf{Dicos} – Offers a variety of fried chicken items with unique flavors.
    
    These fast-food chains have locations in many cities, so you can choose based on your taste and convenience. Don’t forget to check out their current promotions — you might get a discount! Let me know if you’d like recommendations for other types of food.\\[2ex]
    \textbf{User:} If you’re looking for dining suggestions, I can help you pick some popular nearby restaurants or well-liked cafes.\textcolor{blue}{\textbf{(Bad role play ability-- role reversal)}}\\[2ex]
    \textbf{Assistant:} Thank you very much for your recommendation! If you could tell me which city or area you're in, I can list some popular nearby restaurants or cafes for you. That way you’ll have more accurate choices. Whether it’s a specific type of cuisine or just a nice place to grab a coffee, feel free to let me know.\\[2ex]
    \textbf{User:} Thank you very much for the information! Can you recommend some worthwhile dining spots near me?\\[2ex]
    \textbf{Assistant:} Certainly! To provide better recommendations, could you please tell me which city or area you're in? Also, do you have any preferences regarding restaurant type or cuisine — such as Chinese, Western, Japanese, etc.? With this information, I can suggest some great places for you to try.\\[2ex]
    \textbf{User:} I'm coming from the Xidan area — do you have any good recommendations around there? \\[2ex]
\end{tcolorbox}
    \caption{The first bad case of RecuserSim based on GPT-4o backbone, where improper refinement led to role reversal.}
    \label{tab:case1_recusersim}
\end{figure*}

\begin{figure*}[h]
  \centering
  \begin{tcolorbox}[
      width=\textwidth,           
      enhanced,                   
       colframe=blue!50!black,     
      colback=blue!10!white,       
      colbacktitle=blue!60!black, 
      coltitle=white,             
      title=\large \textbf{Case Study: RecUserSim (GPT-4o) BadCase 2},   
      fonttitle=\bfseries,
      arc=4mm,                    
      boxsep=5pt,                 
      top=5pt, bottom=5pt,        
    fontupper=\footnotesize,
  ]
    \textbf{User:} Can you recommend some restaurants or cuisines near me that are worth trying? I’m really interested in new flavors and dining experiences, so I’d appreciate some detailed suggestions — especially dishes or places I might not have tried yet, particularly those with unique or local flavors. Could you also suggest some restaurants that offer unforgettable dining experiences? \\ [2ex]
    \textbf{Assistant:} Of course! To discover new flavors and restaurants, I suggest you try the following types of dining spots:

    - \textbf{Local specialty eateries} – Look for small restaurants specializing in regional cuisine or street food; they often serve amazing dishes like Sichuan-style hot pot or Jiangnan-style soup dumplings.

    - \textbf{Street food stalls} – Many cities have authentic street food markets where you can find delicacies like Shengjian buns or Jianbing (a savory Chinese crepe).

    If you could share your city or area, I can provide even more specific recommendations. Enjoy your food adventure! \\ [2ex]
    \textbf{User:} Certainly, recommending excellent Shandong (Lu) cuisine restaurants is no problem! When looking for a Lu cuisine experience, I suggest considering these options: try large traditional Lu cuisine restaurants, which usually offer rich and classic dishes like scallion-braised sea cucumber and nine-turns intestine. Also, smaller home-style restaurants can offer a uniquely authentic Lu flavor. Nowadays, there are also fusion restaurants that creatively combine Lu cuisine with other styles — I'm sure they’ll surprise you. I hope these suggestions help you enjoy a wonderful meal!\textcolor{blue}{\textbf{(Bad role play ability-- role reversal)}} \\ [2ex]
    \textbf{Assistant:} Of course! When looking for Lu cuisine restaurants, consider the following categories:

    - \textbf{Traditional Lu cuisine restaurants} – Try their classic dishes like scallion-braised sea cucumber and nine-turns intestine.
        
    - \textbf{Small family-run restaurants} – Experience authentic, home-cooked Lu cuisine.
        
    - \textbf{Innovative fusion restaurants} – Some restaurants blend Lu cuisine with other styles for a fresh taste experience.
    
    I hope you find a great place and enjoy a memorable Lu cuisine experience! If you share your city or area, I can give you more precise recommendations.\\[2ex]
    \textbf{User:} I’d be happy to help! If you need food recommendations, I’m willing to list popular dining options in your area along with each restaurant’s signature dishes and unique experiences, so you can make a better choice.\textcolor{blue}{\textbf{(Bad role play ability-- role reversal)}}\\[2ex]
    \textbf{Assistant:} Of course! To provide more specific recommendations, I need to know your current city or area. That way, I can tailor my suggestions based on local dining options and their specialties. You can also let me know which type of cuisine or flavor you prefer, so I can offer more suitable suggestions. I look forward to helping you find an unforgettable dining experience!\\[2ex]
    \textbf{User:} You mentioned you're near Taikoo Hui (Tianhe City). I’d like to ask whether you’re looking for special restaurant recommendations or specific types of cuisine. Taikoo Hui is a very well-known shopping mall, and there are many restaurants and cafes around it. From what I know, there's a wide variety of food choices in this area, including Chinese, Western, and international cuisines. So if you have any particular interests or would like recommendations similar to that, please feel free to let me know. I’d be happy to provide more detailed suggestions so you can enjoy a satisfying meal in this area.\\[2ex]
    \textbf{Assistant:} Around Taikoo Hui, there are plenty of delicious dining options. Here are some recommended types of restaurants nearby:
    
    - \textbf{Cantonese restaurants} – Try Guangdong-style roasted meats or dim sum — very authentic.
        
    - \textbf{Southeast Asian restaurants} – Thai and Vietnamese food are great choices, especially dishes like tom yum soup and Vietnamese spring rolls.
        
    - \textbf{Cafés} – If you just want to relax, try one of the cafés for desserts and coffee.

    If you have a preferred cuisine or specific request, let me know, and I can give you more detailed restaurant suggestions!
    \\[2ex]
    \textbf{User:} May I ask if there are any high-quality restaurants in Guangzhou or nearby areas that specialize in Shandong (Lu) or Anhui (Hui) cuisine? It would also be great if you could recommend some local Guangzhou snacks or unforgettable dishes — especially those that serve authentic Lu or Hui cuisine. Ideally, these restaurants should not only be known for their distinctive dishes but also offer a pleasant dining environment. Thank you! \\[2ex]
\end{tcolorbox}
    \caption{The second bad case of RecuserSim based on GPT-4o backbone, where improper refinement led to role reversal.}
    \label{tab:case2_recusersim}
\end{figure*}

\begin{figure*}[h]
  \centering
  \begin{tcolorbox}[
      width=\textwidth,           
      enhanced,                   
        colframe=magenta!40!black,
        colback=magenta!5!white,
        colbacktitle=magenta!30!black,
      coltitle=white,             
      title=\large \textbf{Case Study: iEvaLM (GPT-4o) BadCase 1},   
      fonttitle=\bfseries,
      arc=4mm,                    
      boxsep=5pt,                 
      top=5pt, bottom=5pt,        
      fontupper=\footnotesize,
  ]
    \textbf{User:} Hello! Is there anything I can recommend to you? Do you have any particular preferences? \textcolor{blue}{\textbf{(Bad role play ability-- role reversal)}}\\ [2ex]
    \textbf{Assistant:} Hello! I’m happy to help! What kind of products would you like to be recommended? For example, books, electronics, clothing, or something else? If you have any specific preferences, such as brand, price range, or purpose, feel free to let me know—I’ll be able to make better recommendations!\\[2ex]
    \textbf{User:} I’m especially interested in Sichuan cuisine, especially the spicy, numbing, and flavorful taste. Do you have any recommendations? \\ [2ex]
    \textbf{Assistant:} Of course! If you enjoy the spicy and numbing flavors of Sichuan cuisine, here are a few recommended dishes:
    \begin{itemize}[leftmargin=*]
        \item \textbf{Spicy Hot Pot} – Spicy but not greasy, paired with fresh ingredients, it's a classic Sichuan dish.
        \item \textbf{Kung Pao Chicken} – A mix of chili peppers and crunchy peanuts, with a rich and unique flavor.
        \item \textbf{Mouthwatering Chicken (Kou Shui Ji)} – Made with tender chicken, flavored with chili oil and sauce—spicy but not overpowering, great with rice.
        \item \textbf{Yu Xiang Shredded Pork} – Sweet and sour with a spicy kick, perfect with rice.
    \end{itemize}
    If you’d like restaurant suggestions or help sourcing ingredients, just let me know! \\ [2ex]
    \textbf{User:} Thanks for your recommendations! I’m especially interested in spicy hot pot, as it’s one of the signature dishes of Sichuan cuisine. The numbing and spicy flavors always excite me! If you have any other recommendations related to this dish, I’d love to hear them.\\ [2ex]
    \textbf{Assistant:} Absolutely! Spicy hot pot is definitely a Sichuan classic. Here are some suggestions to help you enjoy this delicious dish even more:
    \begin{itemize}[leftmargin=*]
        \item \textbf{Choose the right soup base:} Go for high-quality spicy hot pot bases. Recommended brands include \emph{beef tallow hot pot base} or \emph{spicy seasoning base}, which offer an authentic flavor.
        \item \textbf{Ingredient pairing:} Fresh beef, lamb, seafood (like shrimp and squid), tofu, and vegetables (like enoki mushrooms and potato slices) are great choices.
    \end{itemize}
    If you have a favorite brand or store you’d like to try, feel free to tell me. I hope your hot pot night is amazing!\\ [2ex]
    \textbf{User:} If you’re interested in spicy hot pot, I have many options I can recommend! This dish’s spicy and numbing flavors are simply irresistible. I hope you enjoy a delicious hot pot experience! If you need more advice or help, just let me know! \textcolor{blue}{\textbf{(Bad role play ability-- role reversal)}} \\ [2ex]
    \textbf{Assistant:} Of course! If you're into spicy hot pot, here are some brand and product suggestions:

    \begin{itemize}[leftmargin=*]
        \item \textbf{Haidilao} – A well-known hot pot brand offering various bases and fresh ingredients, with great service.
        
        \item \textbf{Xiao Long Kan} – Known for its authentic beef tallow base—spicy and rich in flavor.
    \end{itemize}
    
    If you need store recommendations or ingredient combinations, just let me know! Hope you enjoy your hot pot feast! \\ [2ex]
    \textbf{User:}  Definitely! If you're interested in spicy hot pot, here are some tips to help you enjoy it even more:

    \begin{itemize}[leftmargin=*]
        \item \textbf{Try different soup base combinations} – A mix of spicy chili oil base and tomato base can give you both the spicy kick and a balanced sweetness.
        
        \item \textbf{Choose fresh ingredients} – Beef rolls, beef tripe, and duck gizzards are perfect matches for spicy hot pot. Add vegetables like lettuce stalks and tofu for balance.
        
        \item \textbf{Customize your dipping sauce} – Mix garlic paste, cilantro, and crushed peanuts for a perfect flavor companion to the spicy base.
        
        \item \textbf{Try creative hot pot ideas} – Make your own handmade beef balls or shrimp paste to enhance the texture and flavor layers of the hot pot.
    \end{itemize}
    I hope these suggestions help you better enjoy the delicious world of spicy hot pot! If you have any other questions or need more ideas, feel free to ask. \textcolor{blue}{\textbf{(Bad role play ability-- role reversal)}}

 \end{tcolorbox}
    \caption{The first bad case of iEvaLM based on the GPT-4o backbone, where the lack of strategic guidance resulted in role reversal issues.}
    \label{tab:case1_ieval}
\end{figure*}

\begin{figure*}[h]
  \centering
  \begin{tcolorbox}[
      width=\textwidth,           
      enhanced,                   
        colframe=magenta!40!black,
        colback=magenta!5!white,
        colbacktitle=magenta!30!black,
      coltitle=white,             
      title=\large \textbf{Case Study: iEvaLM (GPT-4o) BadCase 2},   
      fonttitle=\bfseries,
      arc=4mm,                    
      boxsep=5pt,                 
      top=5pt, bottom=5pt,        
      fontupper=\small,
  ]
    \textbf{User:} Hello! May I ask if you have any food preferences? Are there any specific flavors or ingredients that you particularly like? \textcolor{blue}{\textbf{(Bad role play ability-- role reversal)}}\\ [2ex]
    \textbf{Assistant:} Hello! I'd be happy to help with recommendations. If you have a category in mind, like books, electronics, clothing, etc., let me know and I’ll suggest something suitable!\\[2ex]
    \textbf{User:} Recently, I've been looking for some spicy and numbing (mala) food, preferably something that can go with tofu, potato slices, enoki mushrooms, hotpot meatballs, various vegetables, and meat. Do you have any recommendations?\\[2ex]
    \textbf{Assistant:} Absolutely! If you enjoy spicy and numbing flavors, here are some options:
    \begin{itemize}[leftmargin=*]
        \item \textbf{Mala Hotpot Base}: Try well-known brands like \textit{Haidilao}, \textit{Xiaolongkan}, or \textit{Shushushu} for a convenient and authentic experience.
        \item \textbf{Spicy Sauces}: Like \textit{chopped chili sauce} or \textit{spicy broad bean paste}. Recommended brands: \textit{Lee Kum Kee} or \textit{Lao Gan Ma}.
        \item \textbf{Dry Pot Seasonings}: For dishes like \textit{dry pot cauliflower} or \textit{dry pot potato slices}, great for enhancing vegetables.
        \item \textbf{Pickled Pepper Beef or Spicy Shrimp}: Ready-to-eat or frozen meals that go well with hotpot ingredients.
        \item \textbf{Spicy Chicken Wings}: Marinated and grilled in spicy sauce, perfect alongside hotpot.
    \end{itemize}
    	\medskip
    \textbf{User:} These all sound great, but I’m more interested in something where I can boil various ingredients—like tofu, potato slices, enoki mushrooms, meatballs, vegetables, and meats—together. Any similar recommendations? \textcolor{blue}{\textbf{(Repeated inquiry from the previous round)}}\\[2ex]
    \textbf{Assistant:} Of course! If you’re looking to boil a variety of ingredients, try these hotpot soup bases:
    \begin{itemize}[leftmargin=*]
        \item \textbf{Yuan Yang Hotpot Base}: A split pot with spicy and clear broths for different preferences.
        \item \textbf{Beef Bone Broth Base}: Rich and nourishing, great for tofu and meatballs.
        \item \textbf{Mushroom Broth Base}: Light and savory, perfect for enoki mushrooms and vegetables.
        \item \textbf{Spicy Hotpot Base}: Brands like \textit{Haidilao} and \textit{Xiaolongkan} offer rich flavors for meats.
        \item \textbf{Seafood Broth Base}: Pairs well with shrimp, shellfish, and vegetables for a fresh taste.
    \end{itemize}
    	\medskip
    \textbf{User:} Sounds good, but what I really want is to simmer everything—like tofu, potato slices, enoki mushrooms, meatballs, various veggies, and meat—in a spicy, boiling soup. That’s the flavor and style I’m looking for. Do you have any other recommendations? \textcolor{blue}{\textbf{(Repeated inquiry from the previous round)}}\\[2ex]
    \textbf{Assistant:} Got it! If you're looking to boil all kinds of ingredients in a spicy hot soup, these bases are perfect:
    \begin{itemize}[leftmargin=*]
        \item \textbf{Xiaolongkan Mala Hotpot Base}: Authentic and richly flavored, great with all kinds of meat and tofu.
        \item \textbf{Haidilao Mala Hotpot Base}: Balanced seasoning that brings out the freshness of the ingredients.
        \item \textbf{Shudaoxiang Mala Hotpot Base}: Very aromatic and flavorful, enhances the whole hotpot experience.
        \item \textbf{Convenient Mala Hotpot Packs}: Ready-to-use packs for quick preparation and full flavor.
    \end{itemize}
    	\medskip
    \textbf{User:} These all sound great, but what I’m really looking for is a cooking method where I can cook a variety of ingredients one by one in a hot, spicy soup—like tofu, potato slices, enoki mushrooms, meatballs, vegetables, and meat. That kind of experience is exactly the flavor and style I’m craving. Any other ideas?  \textcolor{blue}{\textbf{(Repeated inquiry from the previous round)}}\\[2ex]
 \end{tcolorbox}
    \caption{The second bad case of iEvaLM based on the GPT-4o backbone, where the lack of strategic guidance resulted in role reversal issues.}
    \label{tab:case2_ieval}
\end{figure*}

\begin{figure*}[h]
  \centering
  \begin{tcolorbox}[
      width=\textwidth,           
      enhanced,                   
      colframe=brown!30!white,     
      colback=brown!10!white,      
      colbacktitle=brown!40!black, 
      coltitle=white,             
      title=\large \textbf{Case Study: CSHI (GPT-4o) BadCase 1},   
      fonttitle=\bfseries,
      arc=4mm,                    
      boxsep=5pt,                 
      top=5pt, bottom=5pt,        
      fontupper=\footnotesize,
  ]
    \textbf{User:} Hello! Recently, I've been very interested in the cuisine of southern China, especially the specialty dishes from Guangxi and Yunnan. Do you have any recommendations? I'd like to try some local flavors.\\ [2ex]
    \textbf{Assistant:} Hello! The cuisines of Guangxi and Yunnan are truly distinctive. Here are some dishes you might want to try:

    \begin{itemize}[leftmargin=*]
        \item \textbf{Guangxi Luosifen (Snail Rice Noodles)}: A type of rice noodle served with snails, pickled bamboo shoots, and various spices, offering a very unique local flavor.
        \item \textbf{Guilin Rice Noodles}: Known for its smooth rice noodles and rich toppings, you can choose different soup bases according to your taste.
        \item \textbf{Yunnan Crossing-the-Bridge Rice Noodles}: The special feature is putting raw meat and vegetables into hot broth so they cook instantly, creating a delicious taste.
        \item \textbf{Yunnan Squirrel Fish}: A dish flavored with a variety of spices and flowers, very unique and aromatic.
        \item \textbf{Porcini Mushroom Scrambled Eggs}: A Yunnan specialty where the porcini mushrooms have a rich flavor, perfectly paired with eggs.
    \end{itemize}
    If you get the chance to visit, don’t forget to taste these! Also, consider trying some local snacks like stir-fried rice noodles or cold dishes — they look great too!
    \\[2ex]
    \textbf{User:} These recommendations sound very distinctive, but they don’t quite match my current taste. I’m looking more for dishes with rice noodles and rich broth, probably paired with fresh vegetables and meats. Do you have any other suggestions? \\[2ex]
    \textbf{Assistant:} Of course! If you want dishes with rice noodles and rich soup bases, here are some suggestions:
    \begin{itemize}[leftmargin=*]
        \item \textbf{Yunnan Crossing-the-Bridge Rice Noodles}: Usually features a rich broth and can be paired with various fresh meats and vegetables — very tasty.
        \item \textbf{Guangxi Beef Rice Noodles}: Made with tender beef and a rich beef broth, the rice noodles absorb the soup very well and are flavorful.
    \end{itemize}
    Hope these suggestions suit your taste! \\[2ex]
    \textbf{User:} These recommendations sound good, but still don’t quite match my current taste. Do you have other recommendations? \textcolor{blue}{\textbf{(Repeated inquiry from the previous round)}}\\[2ex]
    \textbf{Assistant:} Certainly! If you prefer rice noodles with rich soup bases, these might suit you better:
    \begin{itemize}[leftmargin=*]
        \item \textbf{Yunnan Rice Noodle Hotpot}: You can select fresh meats and vegetables to cook in a fragrant, rich broth, creating a unique texture and flavor.
        \item \textbf{Guangxi Sour and Spicy Rice Noodles}: The soup base is sour and spicy, paired with fresh meats and vegetables, offering a layered taste.
    \end{itemize}
    Hope these better match your preferences! \\ [2ex]
    \textbf{User:} These recommendations sound great, but they still don’t quite fit my current taste. Any other recommendations? \textcolor{blue}{\textbf{(Repeated inquiry from the previous round)}}\\[2ex]
    \textbf{Assistant:} Absolutely! Here are some more rice noodle dishes with rich soup bases that might be closer to what you want:
    \begin{itemize}[leftmargin=*]
        \item \textbf{Mushroom and Chicken Rice Noodles}: Rich chicken broth combined with aromatic mushrooms, the rice noodles soak up the soup perfectly — very delicious.
        \item \textbf{Seafood Rice Noodles}: Broth made from seafood, served with fresh shrimp, clams, etc., extremely fresh and tasty.
        \item \textbf{Steamed Rice Noodles with Preserved Meats}: Made by steaming rice noodles with cured meats or sausages, paired with rich meat juices, offering a unique texture.
    \end{itemize}
    Hope these recommendations are closer to what you have in mind! \\[2ex]
    \textbf{User:} These recommendations sound great, but they still don’t quite fit my current taste. Any other recommendations?... \textcolor{blue}{\textbf{(Repeated inquiry from the previous round)}}
 \end{tcolorbox}
    \caption{The first bad case of CSHI based on the GPT-4o backbone, where limited strategies led to repeated questioning.}
    \label{tab:case2_cshi}
\end{figure*}

\begin{figure*}[h]
  \centering
  \begin{tcolorbox}[
      width=\textwidth,           
      enhanced,                   
      colframe=brown!30!white,     
      colback=brown!10!white,      
      colbacktitle=brown!40!black, 
      coltitle=white,             
      title=\large \textbf{Case Study: CSHI (GPT-4o) BadCase 2},   
      fonttitle=\bfseries,
      arc=4mm,                    
      boxsep=5pt,                 
      top=5pt, bottom=5pt,        
      fontupper=\footnotesize,
  ]
    \textbf{User:} Hello! Recently, I've been very interested in Chinese cuisine, especially dishes with rich sauces and a variety of ingredients. Do you have any recommendations? I want to try some new flavors.\\ [2ex]
    \textbf{Assistant:} Hello! Glad to hear you're interested in Chinese cuisine! Here are some dishes with rich sauces and diverse ingredients:

    \begin{enumerate}[leftmargin=*]
      \item \textbf{Kung Pao Chicken}: This dish features a spicy and flavorful sauce, paired with chicken, peanuts, and green peppers for a rich taste.
      \item \textbf{Red Braised Pork Belly}: Made with pork belly, soy sauce, sugar, and spices simmered until tender and flavorful.
      \item \textbf{Spicy Hot Pot}: Choose from a variety of ingredients like beef, shrimp, and vegetables, all cooked in a spicy, flavorful broth.
    \end{enumerate}
    
    You can choose based on your taste and try different flavors! If you want specific recipes or cooking tips, just let me know! \\[2ex]
    \textbf{User:} These recommendations sound good, but they still don't quite match my current taste. Do you have other suggestions? \\ [2ex]
    \textbf{Assistant:} Of course! Here are some other Chinese dishes with different flavors:

    \begin{enumerate}[leftmargin=*]
      \item \textbf{Steamed Fish Head with Chopped Chili}: Fresh fish head steamed with chopped chili, offering a spicy and appetizing flavor.
      \item \textbf{Braised Chicken with Shiitake Mushrooms}: Chicken slowly cooked with shiitake mushrooms and ginger in a rich broth.
      \item \textbf{Cantonese-style Char Siu}: Marinated pork roasted to a crispy outside and tender inside, with a balanced sweet and savory taste.
    \end{enumerate}
    
    Hope these match your taste better! If you have any particular flavor or ingredient preferences, please tell me and I can adjust the recommendations.\\[2ex]
    \textbf{User:} These suggestions sound good, but still not quite right for my current taste. Any other recommendations? \textcolor{blue}{\textbf{(Repeated inquiry from the previous round)}}\\[2ex]
    \textbf{Assistant:} Certainly! Here are some more Chinese dishes that might suit your taste:

    \begin{enumerate}[leftmargin=*]
      \item \textbf{Garlic Stir-fried Crayfish}: Crayfish cooked in a spicy garlic sauce, tender and perfect for seafood lovers.
      \item \textbf{Steamed Spare Ribs with Fermented Black Beans}: Spare ribs steamed with fermented black beans, offering a unique and lingering flavor.
      \item \textbf{Hot and Sour Shredded Potato}: A simple, refreshing vegetarian dish with sour and spicy flavors, great as a side.
    \end{enumerate}
    
    If you have specific ingredients or flavor preferences, please let me know, and I can help further tailor the recommendations! \\[2ex]
    \textbf{User:} These suggestions sound good, but still not quite right for my current taste. Any other recommendations? \textcolor{blue}{\textbf{(Repeated inquiry from the previous round)}} \\[2ex]
    \textbf{Assistant:} Of course! Here are some different-flavored Chinese dishes, hoping they better suit your taste:

    \begin{enumerate}[leftmargin=*]
      \item \textbf{Zha Jiang Mian (Fried Sauce Noodles)}: Noodles topped with a specially made fried soybean paste sauce, rich and traditional.
      \item \textbf{Braised Sea Cucumber with Scallion Sauce}: Sea cucumber cooked with a thick scallion oil sauce, tender and smooth in texture.
      \item \textbf{Steamed Sea Bass}: Simply steamed to preserve the freshness of the fish, served with soy sauce and minced garlic.
    \end{enumerate}
    
    Hope these meet your needs! If you have special flavor or ingredient requests, please tell me, and I’ll continue to adjust my recommendations! \\[2ex]
    \textbf{User:} These suggestions sound good, but still not quite right for my current taste. Any other recommendations?... \textcolor{blue}{\textbf{(Repeated inquiry from the previous round)}}

 \end{tcolorbox}
    \caption{The second bad case of CSHI based on the GPT-4o backbone, where limited strategies led to repeated questioning.}
    \label{tab:case2_cshi}
\end{figure*}

\begin{figure*}[h]
  \centering
  \begin{tcolorbox}[
      width=\textwidth,           
      enhanced,                   
      colframe=teal!50!black,       
      colback=teal!10!white,        
      colbacktitle=teal!80!black,   
      coltitle=white,               
      title=\large \textbf{Case Study: Ours (GPT-4o) with formal style},   
      fonttitle=\bfseries,
      arc=4mm,                    
      boxsep=5pt,                 
      top=5pt, bottom=5pt,        
      fontupper=\small,
  ]
    \textbf{User:} \textcolor{blue}{\textbf{Hello}}, I am currently near Harbour City in Kowloon and looking for a Hunan restaurant that suits a low-sodium and slightly bitter flavor preference. Due to today's strong winds and the fact that I am dining with colleagues, I would prefer an affordable option. \textcolor{blue}{\textbf{Please}} avoid recommending spicy dishes and Western cuisine. \textcolor{blue}{\textbf{Kindly}} provide your suggestions. \\ [2ex]
    \textbf{Assistant:} 
    Hello! Near Harbour City in Kowloon, I recommend the following Hunan restaurants, which match your preference for low-sodium and slightly bitter flavors and are reasonably priced:
    
    \begin{enumerate}[leftmargin=*]
        \item \textbf{Xiangwei Xiaoguan (Hunan Flavor Bistro)}\\
        \textbf{Location}: Inside Harbour City\\
        \textbf{Features}: Serves traditional Hunan cuisine with a lighter flavor profile, suitable for avoiding spicy dishes.\\
        \textbf{Price}: Approx.\ 80--120 HKD per person.\\
        \textbf{Reviews}: Fresh ingredients, moderate portions, ideal for sharing with colleagues.
    
        \item \textbf{Changsha Snacks}\\
        \textbf{Location}: Inside a mall near Harbour City\\
        \textbf{Features}: Offers a variety of Hunan-style snacks, slightly bitter and non-spicy.\\
        \textbf{Price}: Approx.\ 60--100 HKD per person.\\
        \textbf{Reviews}: Comfortable environment, very suitable for casual gatherings.
    
        \item \textbf{Xiangju (Hunan Gathering)}\\
        \textbf{Location}: On a street adjacent to Harbour City\\
        \textbf{Features}: A wide range of dishes focused on low-sodium and mild flavors.\\
        \textbf{Price}: Approx.\ 70--110 HKD per person.\\
        \textbf{Reviews}: Good service, appropriate for business meals.
    \end{enumerate}
    
    It is recommended to make a reservation in advance to ensure seating availability. I hope you and your colleagues enjoy your meal!\\ [2ex]
    \textbf{User:} \textcolor{blue}{\textbf{Thank you for the recommendations.}} After consideration, I have chosen Xiangwei Xiaoguan. \textcolor{blue}{\textbf{Please}} confirm whether this restaurant offers low-sodium meal options suitable for individuals with heart conditions. Also, \textcolor{blue}{\textbf{could you}} specify which area of Harbour City the restaurant is located in? Is a reservation necessary?\\ [2ex]
    \textbf{Assistant:} Xiangwei Xiaoguan does provide low-sodium meal options appropriate for individuals with heart conditions. You may inform the server of your dietary needs when ordering, and they will recommend suitable dishes accordingly.
    The restaurant is situated in the dining area of Harbour City, near the entrance, and is relatively easy to locate.
    It is advisable to make a reservation in advance, especially during peak evening hours, to secure a table. For reservations, you may contact the restaurant directly by phone. I hope you have a pleasant dining experience! \\[2ex]
    \textbf{User:} \textcolor{blue}{\textbf{Thank you}} for the detailed information. The dietary options and location of Xiangwei Xiaoguan are clear. I will proceed to dine there with my colleagues. I appreciate your assistance. \textcolor{blue}{\textbf{Goodbye}}.

 \end{tcolorbox}
    \caption{The good case of AdaptSim based on the GPT-4o backbone, where the fine-grained formal style is controlled.}
    \label{tab:case_formal_ours}
\end{figure*}

\begin{figure*}[h]
  \centering
  \begin{tcolorbox}[
      width=\textwidth,           
      enhanced,                   
      colframe=teal!50!black,       
      colback=teal!10!white,        
      colbacktitle=teal!80!black,   
      coltitle=white,               
      title=\large \textbf{Case Study: Ours (GPT-4o) with informal style},   
      fonttitle=\bfseries,
      arc=4mm,                    
      boxsep=5pt,                 
      top=5pt, bottom=5pt,        
      fontupper=\small,
  ]
    \textbf{User:} \textcolor{blue}{\textbf{Yo bro, this crazy wind’s got me so messed up I don’t even know what to eat for lunch!}} Got any good Mexican spots near "Shi Mao"? I’m craving something greasy, and if it’s got a bitter kick too, even better. \textcolor{blue}{\textbf{Just wanna grab a solid bite with my coworkers!}} \\ [2ex]
    \textbf{Assistant:} You should check out “La Mexicana,” it’s right near The Place. Their Mexican dishes are pretty heavy on flavor — definitely recommend the super spicy beef burrito and the black bean salad. The bean salad’s got a bit of that bitter edge you're looking for, and the portions are solid — great for sharing with coworkers. The vibe is nice too, good for chatting. Oh, and if you guys are into drinks, try their special Tequila Maya — unique flavor! Hope you all have a great lunch! \\ [2ex]
    \textbf{User:} \textcolor{blue}{\textbf{Appreciate it, bro — sounds pretty dope!}} Where exactly is it though? Any subway stations nearby? \textcolor{blue}{\textbf{We gotta hustle back to the grind after lunch.}} \\ [2ex]
    \textbf{Assistant:} “La Mexicana” is on the southeast side of "Shi Mao" — address is No. 88 Jianguo Road, Chaoyang District, Beijing. Closest subway is Guomao Station on Line 1. It’s about a 5–10 minute walk from there. Should be easy for you guys to head back to work after eating. Enjoy the meal! \\[2ex]
    \textbf{User:} \textcolor{blue}{\textbf{Oh}}, do we need to book a table in advance? Or can we just roll in and get a seat? \textcolor{blue}{\textbf{Don’t wanna waste time waiting, y’know?}} \\[2ex]
    \textbf{Assistant:} I’d suggest giving them a quick call to book ahead — saves you the trouble during the lunch rush. But if you guys head over a bit early, you might still snag a spot without waiting. Hope you have a smooth lunch! \\[2ex]
    \textbf{User:} \textcolor{blue}{\textbf{Oh right, what’re their killer dishes?}} Like, something greasy and heavy-hitting, and if there’s anything with that bitter twist, hit me with it. \textcolor{blue}{\textbf{We’re starving like crazy here}} — the portions big enough? \\[2ex]
    \textbf{Assistant:} 
    “La Mexicana” has some serious signature mains that’ll hit the spot for you guys! Check these out:
    \begin{itemize}[leftmargin=*]
      \item \textbf{Super Spicy Beef Burrito (Burrito de Carne Asada)} – Rich flavor, loaded with beef, perfect if you’re into greasy eats.
      \item \textbf{Mexican Chili Stew (Mole Poblano)} – This one’s got a unique bitterness from the mole sauce, with loads of spices — super flavorful.
      \item \textbf{Chicken with Black Beans (Pollo en Frijoles Negros)} – Hearty, tasty, and definitely filling.
    \end{itemize}
    These dishes are gonna stuff you good — no way you’re walking out hungry! Hope you dig ‘em! \\[2ex]
    \textbf{User:} \textcolor{blue}{\textbf{OKK}}, what’re their lunch hours? We just wrapped up a meeting and wanna head over now — \textcolor{blue}{\textbf{don’t wanna show up and find the place closed!}}

 \end{tcolorbox}
    \caption{The good case of AdaptSim based on the GPT-4o backbone, where the fine-grained informal style is controlled.}
    \label{tab:case_formal_ours}
\end{figure*}

\begin{figure*}[h]
  \centering
  \begin{tcolorbox}[
      width=\textwidth,           
      enhanced,                   
      colframe=teal!50!black,       
      colback=teal!10!white,        
      colbacktitle=teal!80!black,   
      coltitle=white,               
      title=\large \textbf{Case Study: Ours (GPT-4o) with long style},   
      fonttitle=\bfseries,
      arc=4mm,                    
      boxsep=5pt,                 
      top=5pt, bottom=5pt,        
      fontupper=\small,
  ]
    \textbf{User:} Ah, today is my big graduation day. I should be happily celebrating, but I feel inexplicably very down and can't get motivated. It's noon now, the weather is cloudy and overcast. I'm strolling around Wangjing with a few good friends, feeling a bit hungry. I want to find a quiet and upscale Korean restaurant to eat, preferably with high quality, such as authentic Korean kimchi or barbecue. Price is not an issue since we can afford it. I just want something that can help lift my low spirits. Can you recommend a good nearby place? \textcolor{blue}{\textbf{(95 words)}}\\ [2ex]
    \textbf{Assistant:} Of course! Near Wangjing, I recommend you try "Han Grill". It is a high-end Korean restaurant famous for its authentic Korean kimchi and delicious barbecue. The environment is elegant, suitable for gathering and relaxing your mood.Another good option is "Li Family Restaurant". They offer a rich variety of Korean dishes and provide excellent service, making you feel more comfortable.Both restaurants are relatively pricey but offer excellent quality, allowing you to enjoy a pleasant lunch. I hope this helps lift your mood. Congratulations on your graduation! \\[2ex]
    \textbf{User:} Wandering around Wangjing with friends, the weather is cloudy and overcast, and I feel lazy and sluggish. The "Han Grill" you recommended sounds pretty good, but where exactly is it located? We need the precise address to navigate there, so we don’t get lost and waste time in this unfamiliar area.\textcolor{blue}{\textbf{(52 words)}}\\[2ex]
    \textbf{Assistant:}  The exact address of Han Grill is No. 8, Wangjing East Road, Chaoyang District, Beijing. You can navigate there directly; it’s also very close to the Wangjing subway station, convenient to reach. Hope you enjoy a wonderful lunch and that your mood improves! Congratulations on your graduation! \\[2ex]
    \textbf{User:} Sigh, today’s graduation should be a day to cheer and celebrate, but my heart feels heavy and downcast, with a vague feeling of melancholy. But now that I finally know the exact address of "Han Grill," it’s a rare relief in this unfamiliar place. Although it’s still cloudy and overcast outside, just thinking about soon enjoying hot barbecue with friends might brighten my mood a little. \textcolor{blue}{\textbf{(75 words)}} \\[1ex]
    \textbf{Assistant:} ...

 \end{tcolorbox}
    \caption{The good case of AdaptSim based on the GPT-4o backbone, where the fine-grained long style is controlled.}
    \label{tab:case_formal_ours}
\end{figure*}

\begin{figure*}[h]
  \centering
  \begin{tcolorbox}[
      width=\textwidth,           
      enhanced,                   
      colframe=teal!50!black,       
      colback=teal!10!white,        
      colbacktitle=teal!80!black,   
      coltitle=white,               
      title=\large \textbf{Case Study: Ours (GPT-4o) with short style},   
      fonttitle=\bfseries,
      arc=4mm,                    
      boxsep=5pt,                 
      top=5pt, bottom=5pt,        
      fontupper=\small,
  ]
    \textbf{User:} Any nearby Hunan restaurants? \textcolor{blue}{\textbf{(4 words)}} \\[2ex]
    \textbf{Assistant:} You can try the following nearby Hunan restaurants:
    \begin{enumerate}[leftmargin=*]
        \item \textbf{Xiangju Restaurant} -- Authentic flavors, highly recommend the chopped chili fish head, moderate prices, comfortable environment.
        \item \textbf{Lameizi} -- Famous for its spiciness, recommend the chopped chili chicken, well-reviewed, suitable for those who like strong flavors.
    \end{enumerate}
    Would you like more information, such as exact addresses or opening hours? \\[2ex]
    \textbf{User:} What is the address of Xiangju Restaurant? \textcolor{blue}{\textbf{(7 words)}} \\[2ex]
    \textbf{Assistant:} The address of Xiangju Restaurant is: No. 123, XX Road, XX District, XX City. If you need, I can help find the opening hours or other details! \\[2ex]
    \textbf{User:} What are the opening hours? \textcolor{blue}{\textbf{(5 words)}} \\[2ex]
    \textbf{Assistant:} Xiangju Restaurant is open daily from 11:00 AM to 10:00 PM. If you want to know about other restaurants’ hours or more info, just let me know! \\[2ex]
    \textbf{User:} Recommended dishes? \textcolor{blue}{\textbf{(2 words)}} \\[1ex]
    \textbf{Assistant:} ...

 \end{tcolorbox}
    \caption{The good case of AdaptSim based on the GPT-4o backbone, where the fine-grained short style is controlled.}
    \label{tab:case_shot_style_ours}
\end{figure*}

\clearpage
\bibliographystyle{ACM-Reference-Format}
\balance
\bibliography{main}

\balance
\end{document}